\documentclass{article}

\usepackage[final,nonatbib]{neurips_2022}
\usepackage[numbers]{natbib}
\usepackage{algorithm,algorithmicx,algpseudocode}

\usepackage[utf8]{inputenc} %
\usepackage[T1]{fontenc}    %
\usepackage{hyperref}       %
\usepackage{url}            %
\usepackage{booktabs}       %
\usepackage{amsfonts}       %
\usepackage{nicefrac}       %
\usepackage{microtype}      %
\usepackage{xcolor}         %
\usepackage{graphicx}
\usepackage{multirow}
\usepackage{subfigure}
\usepackage{amsmath}
\usepackage{cleveref}
\usepackage{soul}

\DeclareMathOperator*{\argmin}{arg\,min}

\title{
{Friendly Noise against Adversarial Noise:\\ A Powerful Defense against Data Poisoning Attacks}}

\newcommand{\alg}{\textsc{FrieNDs}}

\newcommand{\figref}[1]{Fig.~\ref{#1}}
\newcommand{\tabref}[1]{Tab.~\ref{#1}}
\newcommand{\secref}[1]{Sec.~\ref{#1}}

\author{%
  Tian Yu Liu \\
  Department of Computer Science\\
  University of California, Los Angeles\\
  \texttt{tianyu@cs.ucla.edu} \\
   \And
  Yu Yang \\
  Department of Computer Science\\
  University of California, Los Angeles\\
  \texttt{yuyang@cs.ucla.edu} \\
   \And
  Baharan Mirzasoleiman \\
  Department of Computer Science\\
  University of California, Los Angeles\\
  \texttt{baharan@cs.ucla.edu} \\ 
}

\begin{document}

\maketitle

\begin{abstract}
A powerful category of (invisible) data poisoning attacks modify a subset of training examples by small adversarial perturbations to change the prediction of certain test-time data. 
Existing defense mechanisms are not desirable to deploy in practice, as they often
either drastically harm the generalization performance, or are attack-specific, %
and prohibitively slow to apply. %
Here, we propose a simple but highly effective approach that unlike existing methods breaks various types of invisible poisoning attacks with the slightest drop in the generalization performance. 
We make the key observation that attacks introduce %
local sharp regions of high training loss, %
which when minimized, results in learning the adversarial perturbations and makes the attack successful.
To break poisoning attacks, our key idea is to alleviate the sharp loss regions introduced by poisons. %
To do so, our approach comprises two components: an optimized friendly noise that is generated to maximally perturb examples without degrading the performance, and a randomly varying noise component.
The combination of both components builds a very light-weight but extremely effective defense against the most powerful triggerless targeted and hidden-trigger backdoor poisoning attacks, including Gradient Matching, Bulls-eye Polytope, and Sleeper Agent. We show that our friendly noise is transferable to other architectures, and adaptive attacks cannot break our defense due to its random noise component. \footnote{Our code can be found at \href{https://github.com/tianyu139/friendly-noise}{https://github.com/tianyu139/friendly-noise}}\looseness=-1
\end{abstract}

\section{Introduction}
Big datasets empower modern over-parameterized deep learning systems. 
Such datasets are often scraped from the internet or other public and user-provided sources.
An adversary can easily insert a subset of malicious examples into the data collected from public sources to harm the model's behavior at test time. 
As a result, deep learning systems trained on public data are extremely vulnerable to data poisoning attacks. Such attacks modify a subset of training examples under small (and potentially invisible) adversarial perturbations, with the aim of changing the model's prediction on specific test-time examples. Powerful attacks generate poisons that visually look innocent and are seemingly properly labeled
\citep{geiping2020witches,huang2020metapoison,souri2021sleeper}. This makes them hard to detect {even} by expert observers.
Hence, data poisoning attacks are arguably one of the most concerning threats to modern deep learning systems \citep{ShankarAML}.

Various types of poisoning attacks have been proposed to challenge and exploit the vulnerabilities of deep learning systems.
Backdoor data poisoning attacks add a fixed but not necessarily visible trigger pattern to a subset of training data as well as the test-time target examples %
\citep{gu2017badnets,souri2021sleeper,turner2018clean}. 
Triggerless poisoning attacks add small bounded perturbations to a subset of training examples to make them similar to the adversarially labeled test-time target in the feature or gradient space %
\citep{aghakhani2021bullseye,geiping2021witches,huang2020metapoison,Shafahi2018poisonfrogs,zhu2019transferable}.
In both cases, training or fine-tuning the model on the poisoned training data causes the model to misclassify certain target examples at test time.

There have been sustained efforts to design effective defense mechanisms \cite{abadi2016deep,chen2019detecting,geiping2021doesn,koh2018stronger,madry2018towards,peri2020deep,shokri2020bypassing,tran2018spectral,yang2022not}. However, existing methods are highly impractical to be employed in real-world deep learning pipelines. Firstly, the majority of the existing methods are attack specific and cannot protect the system against various types of data poisoning attacks \citep{geiping2021doesn,peri2020deep,tran2018spectral}. Secondly, the provided protection is often at the expense of significantly dropping the performance of the machine learning pipeline \citep{abadi2016deep,chen2019detecting,madry2018towards}. Thirdly, existing methods are not effective in protecting the deep learning pipelines against adaptive attacks which can make more powerful poisons with the knowledge of the defense in place \citep{koh2018stronger, shokri2020bypassing}. %
Finally, state-of-the-art defense methods are often so expensive that they can hardly be applied to even medium-sized datasets \citep{geiping2021doesn,peri2020deep}, and are ineffective in presence of larger number of poisons \citep{chen2019detecting,geiping2021doesn,peri2020deep}. 

\begin{figure}[t]
    \centering
    \includegraphics[width=\textwidth]{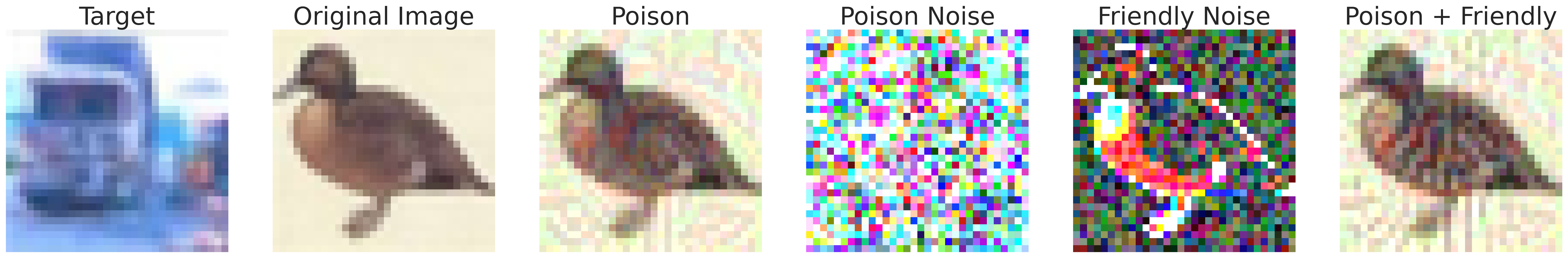}
    \includegraphics[width=\textwidth]{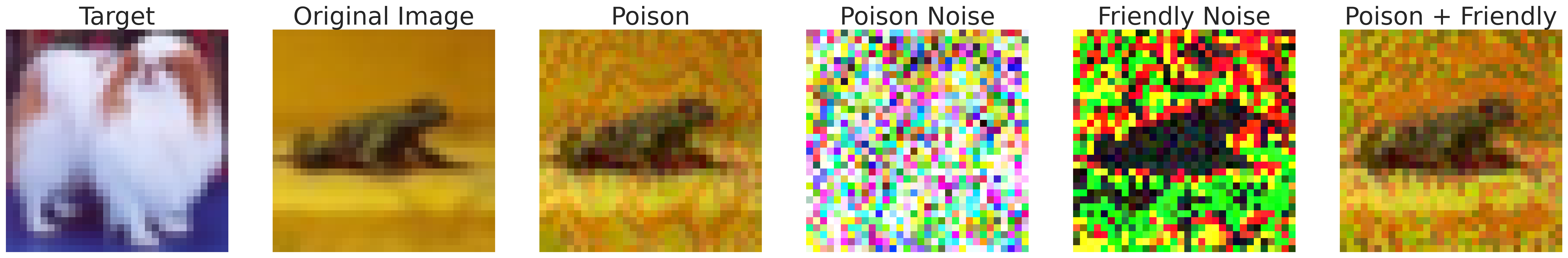}
    \includegraphics[width=\textwidth]{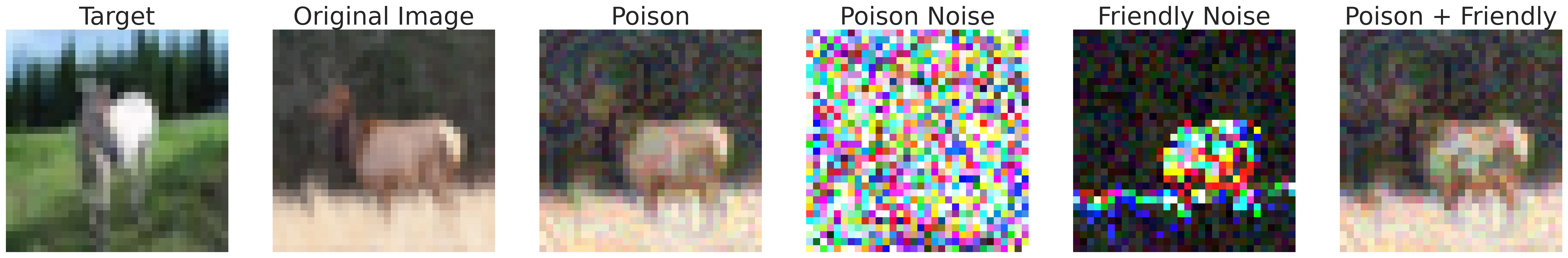}
    \caption{Qualitative Evaluation of Friendly Noise. Our optimized noise adds maximum allowed perturbation to the regions where network robustly learns and leaves other areas untouched (darker regions means less noise). }
    \vspace{-3mm}
    \label{fig:visual}
\end{figure}

In this work, we propose a simple and powerful defense, namely {Friendly Noise Defense} (\alg), against various types of visually imperceptible data poisoning attacks.
In particular, we make the following key observation: data poisoning attacks introduce local sharp regions with high training loss by adding adversarial perturbations to a subset of training examples. 
To effectively break poisoning attacks, our proposed method is composed of two noise components. First, we find the maximum perturbation that can be added to every example without considerably changing the model's output. This fixed accuracy-friendly %
perturbation is found early in training and is transferable to other architectures. Then, we add a small varying random noise in addition to the friendly perturbation to each example at every training iteration. 
Effectively, the two components %
alleviate the local high training loss regions introduced by the poisons, 
and do not allow the attacks to be successful. %
Despite being very lightweight, %
\alg{} can effectively protect deep learning systems against various types of (invisible) poisoning attacks, with a minimum drop in the generalization performance. 

We note that the random noise component of \alg~makes it extremely difficult for an adaptive attacker to break our defense. 
Adaptive attacks can bypass defenses by taking the defense mechanism into account when generating poisons.
For \alg, while an attacker may use the knowledge of the optimization procedure to bypass the friendly noise component, they need to take into account a prohibitively large number of random noise combinations when generating attacks. This makes it extremely difficult for the attacker to ensure the effectiveness of an attack in presence of our \alg{} defense.

Through extensive experiments, we show that our light-weight method renders state-of-the-art visually imperceptible poisoning attacks, including Gradient Matching \cite{geiping2020witches}, Bullseye Polytope \citep{aghakhani2021bullseye}, Feature Collision \citep{Shafahi2018poisonfrogs}, and Sleeper Agent \citep{souri2021sleeper} ineffective, with only a slight decrease in the performance. 
We also show that the optimized noise component generated based on a particular architecture can be applied to defend other architectures against data poisoning attacks. Therefore, it is easy to apply \alg~to real-world deep learning pipelines with minimal additional costs.

\section{Related Work}
\textbf{Targeted Data Poisoning Attacks.}
Data poisoning attacks on deep networks have been explored along two directions - triggered and triggerless attacks. Triggered attacks, or backdoor attacks, aim to misclassify samples containing a `trigger' patch as a pre-determined target class during inference time. In the transfer learning or finetuning setting, earlier works \cite{chen2017targeted,gu2017badnets,liu2017trojaning} relied on label modifications or unbounded image perturbations. 
These attacks are, however, easy to detect.
Subsequently, \cite{Saha2019htbd,souri2021sleeper,turner2018clean} introduced clean-label and visually imperceptible backdoor attacks. Recently, \cite{souri2021sleeper} proposed the first clean-label hidden backdoor attack that is effective on victim models trained from scratch. Triggerless data poisoning attacks aim to misclassify a given target as a pre-determined adversarial class by adding optimized bounded perturbations to a subset of training examples. Such attacks either optimize for feature matching \citep{aghakhani2021bullseye,Shafahi2018poisonfrogs,zhu2019transferable} or %
gradient matching \citep{geiping2020witches} between poisoned and target images, or use meta-learning to solve the poisoning problem directly via bilevel optimization \citep{huang2020metapoison} .

\textbf{Defense Strategies.}
Existing defenses against data poisoning can be divided into filtering and robust training methods. Filtering methods detect outliers in feature space using thresholding \citep{Steinhardt17certified} and nearest neighbors \citep{peri2020deep}, or activation space \citep{chen2019detecting}, or through decomposition of the feature covariance matrix \citep{tran2018spectral}.
These defenses typically assume that only small subsets of the data are poisoned, hence removing such points does not significantly harm generalization. In practice, this assumption may not hold, and such defenses can be easily broken by increasing the number of poisons. Moreover, such methods increase training time by orders of magnitudes, as the filtering step requires %
training the model with poisons, followed by (usually expensive) filtering, and model retraining
\citep{chen2019detecting,peri2020deep,Steinhardt17certified,tran2018spectral}. Very recently, \cite{yang2022not} proposed an efficient method, to iteratively drop examples with isolated (outlier) gradients. %
In comparison, our method is faster, easy to apply, and transferable to other architectures. Hence, it is suitable for deployment in real-world deep learning pipelines.

Robust training methods %
apply randomized smoothing \citep{weber2020rab}, strong data augmentation \citep{borgnia2021strong}, or model ensembling \citep{levine2020deep}. Other methods impose constraints on gradient magnitudes and directions \cite{hong2020effectiveness}, detects and removes poisons with gradient ascent \citep{li2021anti}, or apply adversarial training \citep{geiping2021doesn,madry2018towards,tao2021better}. Deferentially private (DP) training methods have also been explored to defend against data poisoning \citep{abadi2016deep,borgnia2021dp,jayaraman2019evaluating}. Robust training techniques usually involve a significant trade-off between generalization and poison success rate \citep{abadi2016deep,hong2020effectiveness,li2021anti,madry2018towards,tao2021better}, or are computationally very expensive \citep{geiping2021doesn,madry2018towards}. Compared to augmentation-based and adversarial training methods, our method is simple, fast, and maintains good generalization performance. 
Compared to data augmentation, the random noise component of \alg~is considerably more effective in smoothing the loss landscape, due to its much larger space of independent pixel-level transformations.

\textbf{{Random and Adversarial Noise}.}
It is shown that small perturbations can result in large changes in the output of a deep network \cite{szegedy2013intriguing}.
Hence, the application of random and adversarial noise has been studied in various domains. 
In particular, \cite{qin2021random} used Gaussian noise to defend against query-based black box attacks, and 
\cite{rusak2020simple} showed that additive augmentations of Gaussian or Speckle noise is a simple yet very strong baseline for robustness against image corruptions. 
The application of optimized noise has been mainly studied in the context of adversarial training.  
\cite{cai2021learning} used Generative Adversarial Networks (GANs) to generate adversarial perturbations, and \cite{madaan2021learning} relied on meta-learning to learn a noise generator to defend against adversarial perturbations. %
Moreover, \cite{naseer2019cross,xie2019improving} demonstrated the transferability of adversarial perturbations across architectures and domains.
In data poisoning, small random noise generated from a particular distribution has been shown to be ineffective for breaking attacks and harmful to the generalization performance \cite{geiping2021doesn}. %
In contrast, we show that random noise %
combined with our proposed noise optimization approach, 
can make a highly effective defense mechanism against data poisoning attacks and achieve a superior generalization performance.

\begin{figure}[t]\label{fig:match-loss}
\centering
\begin{subfigure}[Matching loss\label{subfig:valley}]{
\includegraphics[width=.28\textwidth]{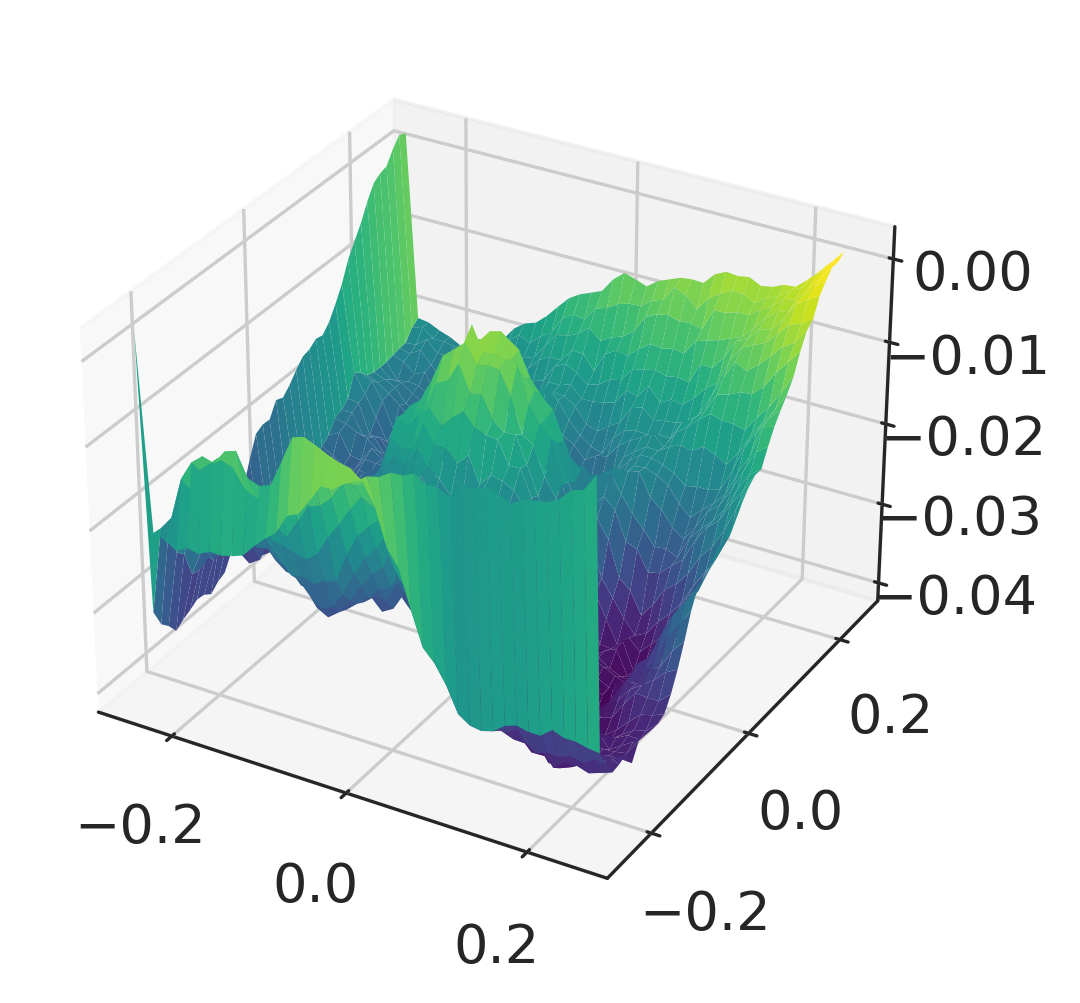}
\vspace{-1mm}
}%
\end{subfigure}
\begin{subfigure}[Matching loss\label{subfig:contour}]{
\includegraphics[width=.36\textwidth]{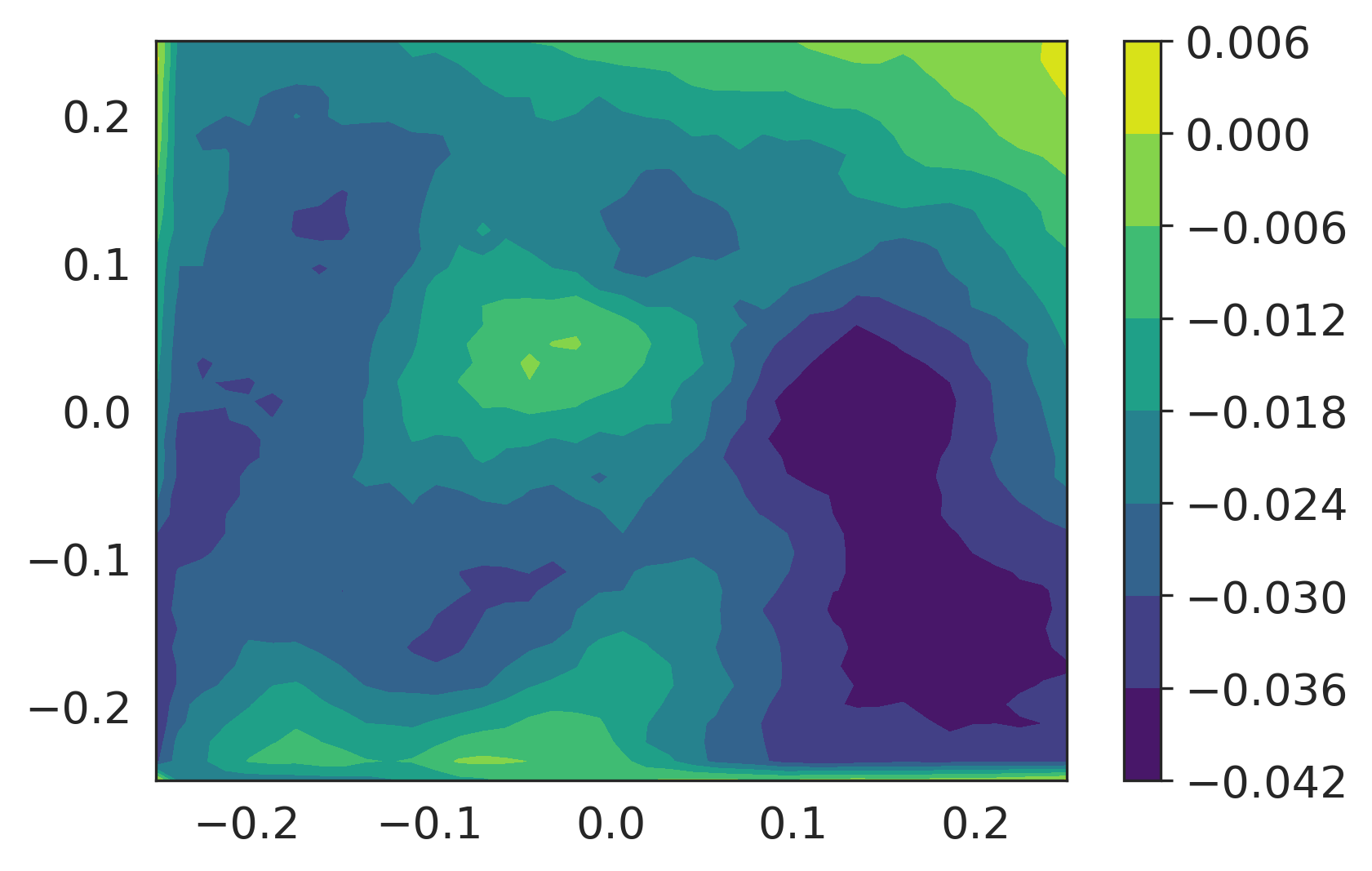}
\vspace{-1mm}
}
\end{subfigure}
\begin{subfigure}[Training loss\label{fig:high-loss-of-single-poison}]{
\includegraphics[width=.28\textwidth]{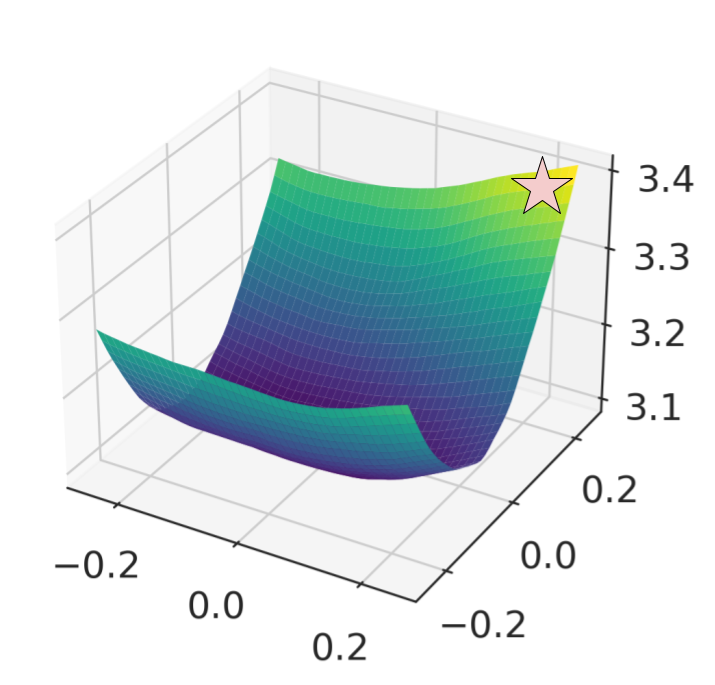}
\vspace{-1mm}
}
\end{subfigure}
\caption{(a) Matching loss defined as $1\!-\!cos\left(\nabla\mathcal{L}(x_t,y_{adv},\theta),\nabla \mathcal{L}(x_i+\delta,y_i,\theta) \right)$, 
as we vary the perturbation $\delta_i$ projected along two randomly chosen directions. (b) Contour view of (a). (c) Training loss of a single example $x_i+\delta_i$ as we vary $\delta_i$, projected along two randomly chosen directions. Adding $\delta^*_i$ which minimizes the matching loss to $x_i$ (located at origin), introduces a local region of high training loss (indicated by a star at $x_i+\delta^*_i$).
}
\end{figure}

\section{\alg: Friendly Noise Defense against Data Poisoning Attacks}

Targeted data poisoning attacks modify a fraction of training data points %
by adding optimized perturbations
that are within an $l_\infty$-norm $\xi$-bound. The optimization is done with the objective of changing the prediction of a target example $x_t$ in the test set, to an adversarial label $y_{\text{adv}}$.
A small perturbation bound $\xi$ ensures that the poisoned examples
remain visually similar to the original (base) training data points. Poisons crafted by such attacks look innocent to human observers and are seemingly labeled correctly. Hence, they are called clean-label attacks.
Targeted clean-label data poisoning attacks can be formulated as the following bilevel optimization problem:
\begin{align}\label{eq:poisoning}
    \min_{\delta \in\mathcal{C}} \mathcal{L}(x_t, y_{\text{adv}}, \theta(\delta)) &\quad s.t.\quad \theta(\delta)\! =\! {\arg\min}_\theta \sum_{i\in V} \mathcal{L}(x_i\!+\!\delta_i, y_i, \theta),
\end{align}
where $\mathcal{C}\!\!=\!\{\delta\!\in\mathbb{R}^{n\times m}\!\!: \!\|\delta\|_\infty\!\!\leq\xi, \delta_i\!=\!0 ~\forall i\notin V_p\}$ is the constraint set defining the set of valid poisons, $V$ is the training data, and $V_p$ is the set of poisoned training examples. 
To address the above optimization problem, powerful poisoning attacks such as Meta Poison (MP) \citep{huang2020metapoison}, Gradient Matching(GM) \cite{geiping2020witches}, Bull-eyes Polytope (BP) \cite{aghakhani2021bullseye}, and Sleeper Agent \citep{souri2021sleeper} craft the poisons to mimic the gradient (equivalently representation in transfer learning) of the adversarially labeled target, i.e.,
\begin{equation}\label{eq:matching}
    \nabla\mathcal{L}(x_t,y_{\text{adv}},\theta)\approx\frac{1}{|V_p|}\sum_{i\in V_p} \nabla\mathcal{L}(x_i+\delta_i,y_i,\theta),
\end{equation} 
Minimizing the training loss on RHS of Eq. \!\!\eqref{eq:poisoning}
also minimizes the adversarial loss on LHS of Eq. \!\! \eqref{eq:poisoning}. 

\subsection{Powerful Poisons {Introduce a Local Sharp Region with High Training Loss}}
Based on Eq. \eqref{eq:matching}, we make the following observation. To substantially change the gradient of a training example $x_i$ to match the adversarial gradient, i.e., $\nabla\mathcal{L}(x_t,y_{\text{adv}},\theta)\approx \nabla\mathcal{L}(x_i+\delta_i,y_i,\theta)$, under bounded perturbation %
$\|\delta_i\|_\infty\leq\xi$, 
the attacker needs to 
exploit the highly non-convex nature of the loss. That is, the attacker needs to find 
regions in a ball of radius $\xi$ around example $x_i$, for which $\nabla \mathcal{L}(x_i+\delta_i,y_i,\theta)$ is considerably different than $\mathcal{L}(x_i,y_i,\theta)$.
Fig. \ref{subfig:valley}, \ref{subfig:contour} illustrate the \textit{matching loss} %
between the gradient of the adversarial loss and the gradient of a perturbed training example, as we vary the perturbation $\delta_i$ projected along two randomly chosen directions. The matching loss is defined as $1\!-\!cos\left(\nabla\mathcal{L}(x_t,y_{adv},\theta),\nabla \mathcal{L}(x_i+\delta,y_i,\theta) \right)$, %
where $cos(u,v)$ is the cosine similarity between vectors $u$ and $v$. 
We see that the adversarial perturbation $\delta_i$ can be effectively optimized to minimize the matching loss in the darker valleys around $x_i$. %
Such valleys do not exist %
around all the examples, and hence not every example can be perturbed in a ball of radius $\xi$ to match the adversarial gradient. Indeed, the examples that can be perturbed by $\delta_i$ s.t. $\|\delta_i\|\leq\xi$ to closely match the adversarial gradient are \textit{effective poisons} \cite{yang2022not}, that make the attack successful.
Crucially, each effective poison %
introduces a %
local increase to the {training loss}, as demonstrated by Fig. \ref{fig:high-loss-of-single-poison}.
The set of effective poisons together introduce a \textit{local sharp region} with a considerably high \textit{training loss}, as illustrated by Fig \ref{fig:loss-landscape-nodef}.
Minimizing the training loss on the poisoned data results in learning the adversarial perturbations, and hence a successful attack.

As poisons %
need to match a particular gradient or representation, they are highly sensitive to small perturbations. 
In \figref{fig:noise_deviations} in the Appendix, we show that the model output has indeed a very high standard deviation around the poisoned examples.
Therefore, slightly perturbing the poisons %
considerably changes their gradient and make them ineffective. %
The main idea behind our friendly noise defense method, \alg, is to {maximally perturb the training examples to make the effective poisons ineffective.}
However, perturbing the training examples should be done in a way that does not harm the generalization performance of the model.
To address this, our method is composed of two components: First, we find the maximum perturbation that can be added to every training example without changing its prediction. 
This alleviates the local sharp region of high training loss introduced by the effective poisons.
To further break the attack, we also add a small varying random noise to every example during the training. %
The random noise generally smooths out the training loss and further %
alleviates the local sharp regions introduced by effective poisons. Thus, both components together can effectively break the attacks. Below, we discuss each component in more details.

\subsection{Optimizing the Friendly Noise: Maximally Perturbing Examples without Harm}

\begin{figure}[t]
\centering
\begin{subfigure}[Undefended\label{fig:loss-landscape-nodef}]{
 \includegraphics[width=.3\textwidth]{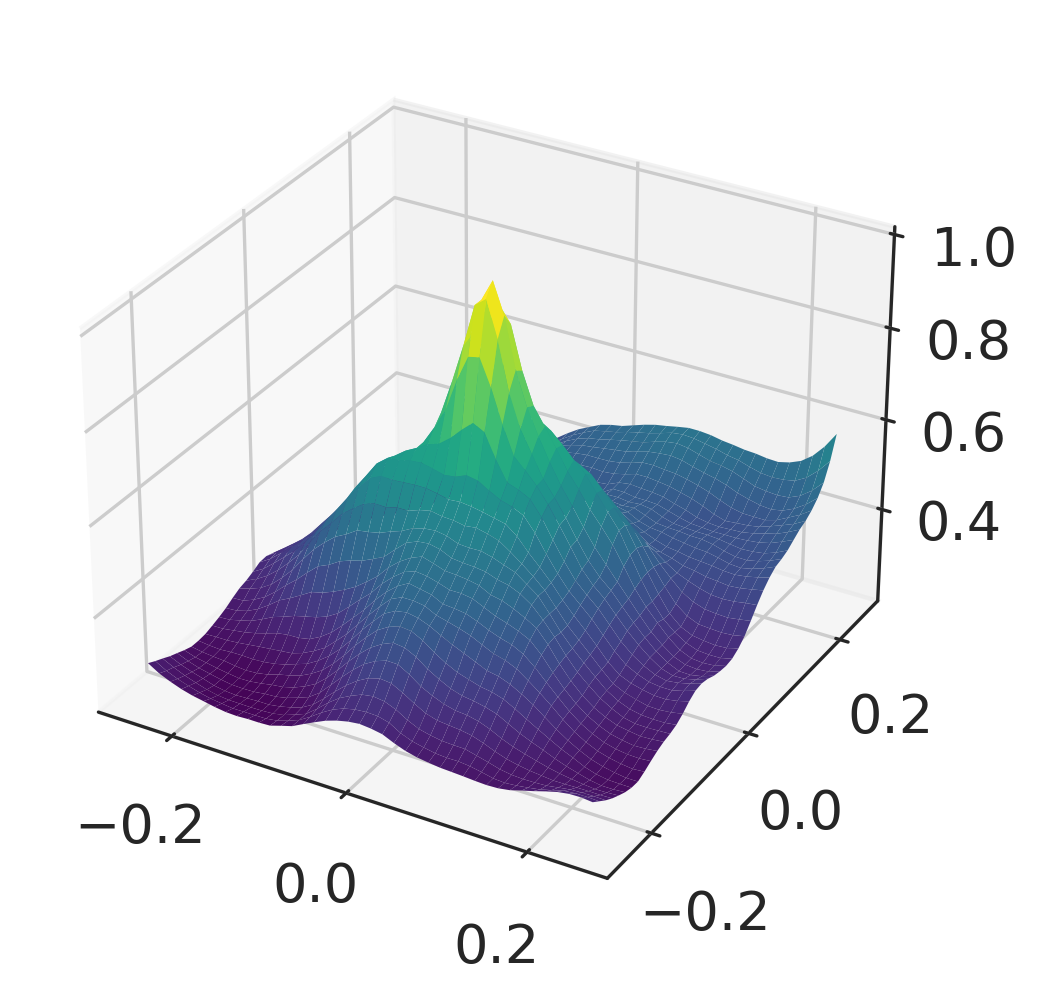}
 \vspace{-1mm}
}
\end{subfigure}
\begin{subfigure}[Friendly noise\label{fig:loss-landscape-friends}]{
 \includegraphics[width=.3\textwidth]{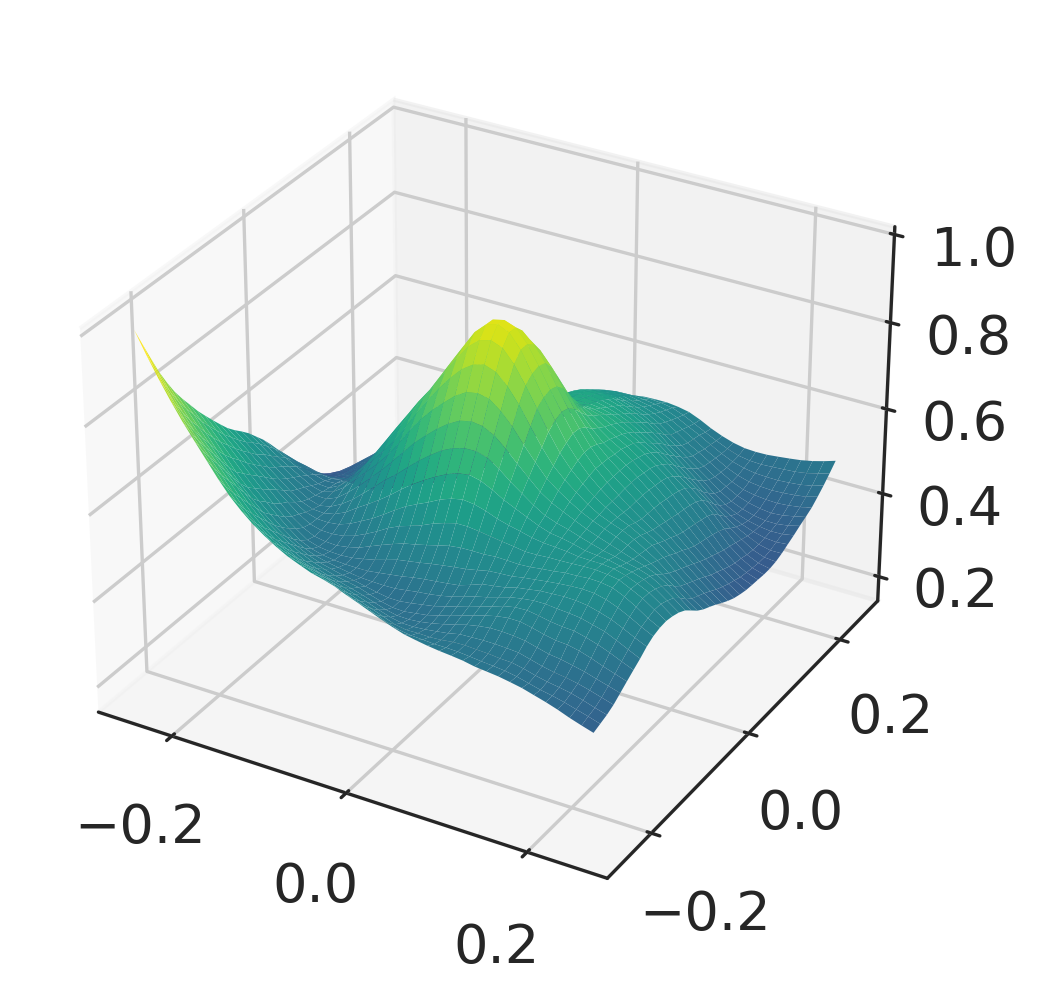}
 \vspace{-1mm}
}
\end{subfigure}
\begin{subfigure}[Random noise\label{fig:loss-landscape-bernoulli}]{
 \includegraphics[width=.3\textwidth]{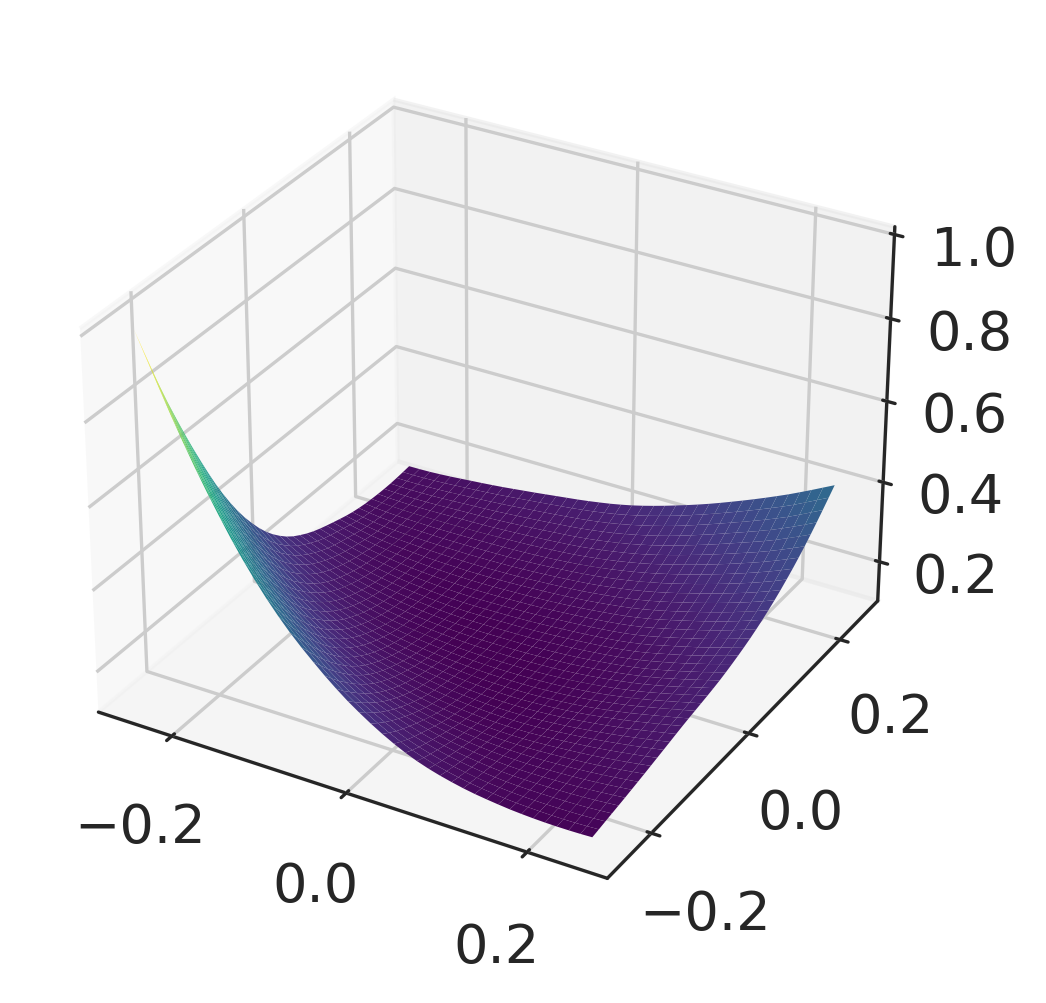}
 \vspace{-1mm}
}
\end{subfigure}
\caption{(a) Loss landscape of a victim model around \textit{all} poisoned examples. %
Effective poisons introduce a local sharp region with high training loss. (b) Training loss landscape of a victim model defended with friendly noise. (c) Training loss landscape of a victim model defended with random Bernoulli noise. %
Both components (\alg) smooth the local sharp region introduced by poisons to the training loss (a) and
reduce the effectiveness of the attack.\looseness=-1
}
\vspace{-1mm}
\end{figure}

The first component of our method finds the maximum perturbation that can be added to every example without significantly changing the model's output.
To do so, for every example $x_i$ we optimize for the largest noise $\epsilon_i$ within an $l_\infty$ norm $\zeta$-bound that results in similar prediction probabilities, measured by KL-divergence.
Formally for each example $x_i$, we find perturbation $\epsilon_i$ as follows\footnote{We show in Appendix \ref{app:norm-ablation} that using ${\|\epsilon\|}_1$ or ${\|\epsilon\|}_\infty$ in Problem \eqref{eqn:noise-objective} yields similar results.}:\looseness=-1
\begin{align}
\label{eqn:noise-objective}
    \epsilon_i = \argmin_{\epsilon:\|\epsilon\|_\infty\leq\zeta} {D_{KL}\big(f_\theta(x_i+\epsilon) || f_\theta(x_i)\big)} - \lambda {\|\epsilon\|}_2. %
\end{align}
We generate a fixed accuracy-friendly perturbation for every data point by solving problem \eqref{eqn:noise-objective} once, using a few Stochastic Gradient Descent (SGD) steps.
There is a trade-off between the time of generating the optimized perturbations and their effectiveness. %
In particular, the optimized perturbations need to be generated and added to the training examples early in training, before %
the attack succeeds (note that we aim to prevent the attacks from being successful). 
At the same time, 
for the perturbations to be effective, they should be generated after the decision boundary is shaped, so that they can be optimized to minimize the change in the model and its decision boundary. {We visualize this trade-off in Appendix \ref{app:friends-begin-time-tradeoff}}. 
We found that training for as few as 5 epochs before solving the optimization problem \eqref{eqn:noise-objective} yields effective perturbations that reduce the attack success rate without harming the model's performance.
The pseudocode can be found in Alg.\ref{alg:generate-noise}.

To better understand the effect of our friendly noise, we illustrate the histogram of the noise added to every pixel in the training data in Appendix \ref{app:histogram-friends},
\figref{fig:histogram-friendly-noise}.
We see that our method mainly targets certain pixels in every image by adding the maximum amount of perturbation, and leaves the rest of the image untouched. 
Certain semantic regions have been shown to be much more robust to perturbations \citep{berger2021stereoscopic}. 
Perturbing the more robust areas does not considerably change the model's behavior and training dynamics. On the other hand, the least robust areas are very sensitive to perturbations, hence small amounts of noise in such areas result in a relatively large change in the model's behavior. %
We visualize the poisons before and after adding our friendly noise in Fig. \ref{fig:visual}. 
We see that our friendly noise successfully targets certain areas in every image that are robustly learned by adding the maximum possible perturbation.

The maximal friendly perturbation added to every example by our method prevents the effective poisons to closely match the target gradient or representation.
Effectively, this alleviates the local sharp region of high training loss introduced by the effective poisons, as is illustrated in Fig. \ref{fig:loss-landscape-friends}, and 
considerably reduces the success rate of the attack.
At the same time, the important features used for classifications are preserved and hence model predictions remain unchanged.
Hence, our method reduces the attack success rate while %
ensuring only a slight drop in the test accuracy. %

Next, we discuss how adding a random variable random noise %
can further improve the model's robustness against poisoning attacks. %

\subsection{Adding Random Noise: Further Smoothing the Loss}

\begin{algorithm}[tb]
   \caption{Generating Friendly Noise}
   \label{alg:generate-noise}
\begin{algorithmic}
   \Require{Train dataset $X$, Model $f_\theta$, LR $\eta_{opt}$, $\lambda$, small number $T$}
   \For{$i \in [T]$}
   \State  $\theta^i = \theta^{i-1} - \eta \nabla_{\theta} L(\theta^{i-1}, {X})$ \Comment{Train the model for a few
   epochs}
   \EndFor
   \For{$x_i \in X$}
   \State Initialize noise $\epsilon_i^0$ uniformly sampled from $[-\epsilon_{init} , \epsilon_{init}]$
   \For{$t=1$ to $T$}
   \State $\epsilon_i^t = \epsilon_i^{t-1} - \eta_{opt} \nabla_\epsilon ({D_{KL}\left(f_\theta(x_i+\epsilon_i^{t-1}) || f_\theta(x_i)\right)} - \lambda {\|\epsilon_i^{t-1}\|}_2)$
   \EndFor
   \State Store noise $\epsilon_i = \epsilon_i^T$ for example $x_i$
   \EndFor
\end{algorithmic}
\end{algorithm}

\begin{algorithm}[tb]
  \caption{Training with \alg{}}
  \label{alg:training}
\begin{algorithmic}
  \Require{Train dataset $X$, Random Noise Distribution A, Epoch to start defense $def\_epoch$}
  \State Run Algorithm 1 to generate $\{\epsilon_i\}_{i=1}^{|X|}$ %
  \For{$i=def\_epoch$ to $n\_epochs$}
  \For{$x_i \in X$}
  \State Sample random noise $\mu_i \sim A$
  \State Set $\hat{x}_i = x_i + \epsilon_i + \mu_i$ and add to dataset $\hat{X}$
  \EndFor
  \State $\theta^i = \theta^{i-1} - \eta \nabla_{\theta} L(\theta^{i-1}, \hat{X})$ \Comment{SGD update step with new dataset $\hat{X}$} 
  \EndFor
\end{algorithmic}
\end{algorithm}

As discussed, our friendly perturbation mainly targets the robust areas of the image and alleviates the local sharp region with high training loss introduced by effective poisons, %
without considerably changing the training dynamics. However, it does not affect  adversarial perturbations that are added to the areas that are not robustly learned.
To further improve the model's robustness against poisoning attacks, we add a small variable random noise to all the training examples at every training iteration. %
Adding the variable random noise generally smooths out the loss landscape. In doing so, it further alleviates the local sharp regions introduced by the attack,
and further drops the attack success rate.
Note that adding a large random noise can harm the test accuracy by over-smoothing the loss. However, the combination of friendly noise and a smaller random noise can effectively break poisoning attacks without harming the model's performance.
\figref{fig:loss-landscape-friends} and \figref{fig:loss-landscape-bernoulli} show that
the local region of high training loss introduced by effective poisons is alleviated
after adding each component of our defense.  This clearly demonstrates the effect of our defense and explains its effectiveness. 
The pseudocode of our defense, \alg, can be found in Alg. \ref{alg:training}.

The small varying random noise can be sampled from various distributions. In particular, we sample from 3 different random noise distributions: (1) Bernoulli: noise is randomly sampled from $\{-\mu, \mu\}$, (2) Uniform: noise is randomly sampled from $[-\mu, \mu]$, and (3) Gaussian: noise is sampled from the normal distribution $\mathcal{N}(0, \mu)$. We note that compared to Uniform noise, Bernoulli and Gaussian noise are in general more effective at reducing the attack success rate, as they make larger perturbations to
input features. At the same time, this makes %
them to suffer a larger drop in the test accuracy. We visualize and compare the Bernoulli, Uniform, and Gaussian noise types in Appendix \ref{app:histogram-friends}.

\subsection{Adaptive attacks}
\label{sec:adaptive-attacks}
Adaptive attacks can respond to a novel defense algorithm when the attacker is aware of the defense. 
If the defense algorithm is known to the attacker beforehand, the attacker can generate more powerful poisons by taking into account the specific defense in place. For example, Gradient Matching \cite{geiping2020witches} and Sleeper Agent \cite{souri2021sleeper} demonstrated that by including augmented examples as well as original examples during poison generation in Eq. \eqref{eq:matching}, they can obtain
robustness against standard data augmentations like crops and flips, when the augmentation technique is preempted by the attacker. 
For \alg, the prohibitively large search space of random noise permutations and its pixel-wise independence property make it extremely difficult for adaptive attacks to break. 
That is, while an attacker may use the knowledge of the time and optimization procedure to bypass the friendly perturbation component of \alg, they need to take into account a prohibitively large number of random noise combinations to bypass the random noise component. 
For example, for a fixed Bernoulli noise, with $p$ pixels there are $2^p$ combinations that an attacker should take into account to ensure the poisons' effectiveness. For real images, this becomes prohibitively expensive. Similarly, for a fixed Gaussian and uniform noise, an infinite number of combinations must be considered during the poison generation to ensure the attack's robustness. Note that our method applies a varying random noise at every iteration, which also needs to be taken into account by the attacker.
This makes \alg~robust against adaptive attacks. %

\section{Experiments}

\begin{table}[t]
    \centering
    \caption{Baselines - Against Gradient Matching eps=16, 80 epochs. For trials with all equal outcomes, we report worst-case error estimate $5.59\%$}
    \begin{tabular}{c|ccc}
    \toprule
    Defense & Poison Acc & Test Acc & Time (HH:MM) \\  
    \midrule
    AP-0.25 \cite{geiping2021doesn} & $15.00\% \ (\pm 2.85\%)$ & $93.27\%\ (\pm 0.00\%)$ & 02:39 \\
    AP-0.5 \cite{geiping2021doesn}& $10.00\% \ (\pm 2.01\%)$ & $92.83\%\ (\pm 0.00\%)$ & 03:41 \\
    \textbf{AP-0.75} \cite{geiping2021doesn} & $\mathbf{0.00\% \ (\pm 5.59\%)}$ & $\mathbf{91.29\%\ (\pm 0.00\%)}$ & 04:30 \\
    DeepKNN \cite{peri2020deep} & $75.00\%\ (\pm 4.19\%)$ & $93.72\%\ (\pm 0.24\%)$ & 02:55 \\
    Adversarial Training \cite{madry2018towards} & $60.00\%\ (\pm 5.37\%)$ & $92.03\%\ (\pm 0.31\%)$  & 02:37 \\
    Activation Clustering \cite{chen2019detecting} & $45.00\%\ (\pm 5.53\%)$ & $87.69\%\ (\pm 0.50\%)$ & 01:01\\
    Diff. Priv. SGD \cite{hong2020effectiveness} & $5.00\%\ (\pm 1.06\%)$ & $75.70\%\ (\pm 1.19\%)$ & 00:38 \\
    \textbf{EPIC-0.1} \cite{yang2022not} & $\mathbf{0.00\% \ (\pm 5.59\%)}$ & $90.04\%\ (\pm 0.22\%)$ & \textbf{00:48} \\
    Friendly Noise & $10.00\% \ (\pm 2.01\%)$ & $91.73\%\ (\pm 0.30\%)$ & \textbf{00:37} \\
    \alg -U & $5.00\% \ (\pm 1.06\%)$ & $91.91\%\ (\pm 0.28\%)$ & \textbf{00:37} \\
    \textbf{\alg -B} & $\mathbf{0.00\% \ (\pm 5.59\%)}$ & $\mathbf{91.52\%\ (\pm 0.28\%)}$ & \textbf{00:37} \\
    \textbf{\alg -G} & $\mathbf{0.00\% \ (\pm 5.59\%)}$ & $\mathbf{91.50\%\ (\pm 0.25\%)}$ & \textbf{00:37} \\
    \bottomrule
    \end{tabular}
    \label{tab:wb16}
\end{table}

\begin{table}[t]
    \centering
    \caption{Comparisons with Sleeper Agent defenses averaged over 24 datasets (40 epoch setting). ($\ast$) To accommodate original two-stage learning setup, we ran this for 85 epochs.}
    \begin{tabular}{c|cc}
    \toprule
         Defense & Poison Acc & Test Acc \\
         \midrule
         None & $83.48\%\ (\pm 7.58\%)$ & $91.56\%\ (\pm 0.19\%)$ \\
         Spectral Signatures \cite{tran2018spectral} & $37.17\%\ (\pm 10.10\%$) & $89.94\%\ (\pm 0.19\%)$ \\
         Activation Clustering \cite{chen2019detecting} & $15.17\%\ (\pm 5.38\%)$ & $72.38\%\ (\pm 0.48\%)$ \\
         Diff. Priv. SGD \cite{hong2020effectiveness} & $13.14\%\ (\pm 4.49\%$) & $70.00\%\ (\pm 0.17\%$) \\
         Strong Augmentation \cite{borgnia2021dp} & $69.75\%\ (\pm 10.77\%)$ & $91.32\%\ (\pm0.12\%)$ \\
         STRIP \cite{gao2019strip} & $62.68\%\ (\pm 4.90\%)$ & $92.23\%\ (\pm 0.05\%)$ \\
         NeuralCleanse \cite{wang2019neural} & $85.11\%\ (\pm 5.04\%)$ & $92.26\%\ (\pm 0.06\%)$ \\
         ABL \cite{li2021anti} ($\ast$) & $67.21\%\ (\pm 10.52\%)$ & $81.33\%\ (\pm 3.21\%)$\\
         EPIC-0.1 \cite{yang2022not} & $23.93\%\ (\pm 3.48\%)$ & $89.56\%\ (\pm 0.14\%)$ \\
         EPIC-0.2 \cite{yang2022not} & $\mathbf{9.10\%\ (\pm 1.57\%)}$ & $86.21\%\ (\pm 0.14\%)$ \\
         \textbf{\alg-B} & $\mathbf{21.52\%\ (\pm 8.39\%)}$ & $\mathbf{89.76\%\ (\pm 0.30\%)}$ \\
         \textbf{\alg-G} & $\mathbf{22.99\%\ (\pm 8.59\%)}$ & $\mathbf{89.87\%\ (\pm 0.31\%)}$ \\
         \alg-U & $34.53\%\ (\pm 9.71\%)$  &  $90.36\%\ (\pm 0.38\%)$  \\
         \bottomrule
    \end{tabular}
    \label{tab:sleeper}
\end{table}

\begin{table}[t]
    \centering
    \caption{Against different data poisoning attacks. Here, we use \alg-B as the defense method. Note: Baseline metapoison is ran without default augmentations, following settings used in \citep{geiping2021doesn,huang2020metapoison}}
    \begin{tabular}{c|c|cc|cc}
    \toprule
    & & \multicolumn{2}{c|}{Undefended} & \multicolumn{2}{c}{Defended} \\
    Attack & Scenario & Posion Acc & Test Acc & Poison Acc & Test Acc \\  
    \midrule
    Gradient Matching ($\xi=8$) & From-scratch & 50.00\% & 93.55\% & 0.00\% & 91.55\%  \\
    Gradient Matching ($\xi=16$) & From-scratch & 75.00\% & 93.50\% & 0.00\% & 91.52\% \\
    Metapoison ($\xi=8$) & From-scratch & 45.00\% & 87.61\% & 20.00\% & 90.82\% \\
    Bullseye Polytope & Transfer & 100.00\% & 92.13\% & 35.00\% & 79.35\% \\
    Poison Frogs & Transfer & 100.00\% & 92.12\% & 30.00\% & 79.07\% \\
    Sleeper Agent & From-scratch & 91.72\% & 93.36\% & 31.20\% & 91.31\% \\
    \bottomrule
    \end{tabular}
    \label{tab:attacks}
\end{table}

\begin{table}[h]
    \centering
    \footnotesize
    \caption{Ablation study on random noise components of \alg{} using Gradient Matching attack ($\xi=16$). We set $\zeta=32$ for Friendly Noise, and $\mu=32$ for Noise Only. For experiments on \alg{}, we set $\zeta=16$, $\mu=16$ to combine each component proportionately.}
    \begin{tabular}{cc|cc|c|cc|cc}
    \toprule
     \multicolumn{2}{c|}{No Def.} & \multicolumn{2}{c|}{Friendly Noise} & \multirow{2}{*}{Noise Type} & \multicolumn{2}{c}{Noise Only} & \multicolumn{2}{c}{\alg} \\
    \cmidrule(lr){1-2} \cmidrule(lr){3-4} \cmidrule(lr){6-7} \cmidrule(lr){8-9}
    P. Acc. & Test Acc. & P. Acc. & Test Acc. & & P. Acc. & Test Acc. & P. Acc. & Test Acc. \\
    \midrule
    & & & & Gaussian & \textbf{0.00} & 89.46 & \textbf{0.00} & \textbf{91.50} \\
    75.00 & 93.50 & 10.00 & 91.73 & Bernoulli & 5.00 & 89.31 & \textbf{0.00} & \textbf{91.52} \\
     & & & & Uniform & \textbf{0.00} & 91.61 & 5.00 & \textbf{91.91}\\
    \bottomrule
    \end{tabular}
    \label{tab:noise}
\end{table}

\subsection{Implementation details}

We evaluate our defense method against both triggerless data poisoning and backdoor attacks under two attack settings - training from scratch and transfer learning. 
Following the works of \citep{geiping2020witches,geiping2021doesn,schwarzschild2020just}, we evaluate our method primarily on CIFAR-10, ResNet-18. We also normalize and augment training images with default CIFAR-10 augmentations as used in \cite{geiping2021doesn}. 
For all models trained from scratch, we use a learning rate starting at 0.1 and decaying by a factor of 10 at epochs 30, 50, and 70. For transfer learning, we decay the learning rate at epochs 15, 25, and 35.
When applying our method, we clamp the generated friendly perturbations using $\zeta=16$, and add bounded random noise. For the random noise component, we set $\mu=16$ in our experiments. We also normalize the image as a pre-processing step. We optimize friendly perturbations using SGD with momentum 0.9 and Nesterov acceleration,  perform a hyperparameter search along LR$=\{10,20,50,100\}$ and $\lambda=\{1,10\}$, and optimize each batch of 128 samples for 20 epochs. Following previous works, we report poison success rate (or poison accuracy) as the percentage of datasets poisoned at the end of training. We run all experiments and timings on an NVIDIA A40 GPU.

\subsection{From-Scratch Setting}
First, we evaluate our method on poisoning attacks targeted toward victim models trained from scratch. Such an attack assumes a gray-box scenario, where attackers have knowledge of the victim architecture, but have no knowledge of the specific initialization of the victim's model. Similar to the settings used in \cite{schwarzschild2020just}, which proposes a standardized benchmark for backdoor and data poisoning attacks, benchmark settings, we generate poisoning attacks by selecting $1\%$ of training examples as poisons, which are perturbed within the $l_\infty$ ball of some radius $\xi$. Unless otherwise specified, we set $\xi=16$. The victim model is initialized with the same architecture targeted by the attack based on a different random seed, and is trained on the poisoned dataset using SGD. When applying \alg{}, we set $def\_epoch=5$, and train only with random noise for the first 5 epochs.

\subsubsection{Baseline Comparison and Ablation Study}
We evaluate our method and baseline defenses against the Witches' Brew, or Gradient Matching, attack \citep{geiping2020witches}. It is the current state-of-the-art among data poisoning attacks when applied to the from-scratch setting, and is adapted to be effective against data augmentation and differential privacy \citep{geiping2021doesn}. 
We follow the settings proposed by \cite{geiping2021doesn}, under which we generate 20 different attack datasets for ResNet-18 trained on CIFAR10 with a $1\%$ budget bounded by $\xi = 16$, with a slight modification - while \cite{geiping2021doesn} uses 40 epochs for training, we use 80 epochs to show that our method easily scales to real-world training pipelines. This is because 40 epochs of training only yields $92.01\%$ test error, while 80 epochs yield $93.50\%$.
In \tabref{tab:wb16}, we show that we outperform state-of-the-art defenses \cite{chen2019detecting,geiping2021doesn,hong2020effectiveness,madry2018towards,peri2020deep,yang2022not}. Notably, we achieve the same $0.00\%$ poison success rate with $91.52\%$ test accuracy, an improvement over of $0.23\%$ over state-of-the-art \cite{geiping2021doesn} which yield $91.29\%$ test accuracy at the same poison success rate. Most importantly, \alg{} completes in 37 mins, $7.3$x faster than \cite{geiping2021doesn} which completes in 4.5hrs. 
Compared to the efficient method of \cite{yang2022not}, namely EPIC-0.1 {(with $T=1, K=5$)}, \alg{} achieves a $1.48\%$ higher accuracy while being 1.2x faster. We also strongly outperform other baseline defense methods simultaneously in all three metrics - poison success rate, test accuracy, and runtime. In \tabref{tab:attacks}, we show that \alg{} also effectively defends against MetaPoison \cite{huang2020metapoison}, reducing the poison success rate from $45.00\%$ to $20.00\%$ with an accuracy gain from $87.61\%$ to $90.82\%$ resulted from applying augmentations.

We further show that our approach is effective against backdoor attacks, in particular, against the Sleeper Agent attack \cite{souri2021sleeper}. Sleeper Agent is the current state-of-the-art clean-label backdoor attack, and the only such attack shown to be effective in from-scratch settings. Following their evaluation protocol, we generate 24 poisoned datasets with $\xi=16$, and evaluate our defense by training 24 victim models respectively for 40 epochs and testing the poison success rate on 1000 target backdoor images per dataset. We compare our method against other defenses evaluated by \citep{souri2021sleeper} in \tabref{tab:sleeper}. Here, \alg{} successfully defends against \cite{souri2021sleeper} by reducing poison accuracy from $83.48\%$ to $21.52\%$ with only a small drop in test accuracy from $91.56\%$ to $89.76\%$. We outperform the next best methods, 
EPIC-0.1 \cite{yang2022not} {(with $T=2, K=5$)} and Spectral Signatures \cite{tran2018spectral}, by lowering poison accuracy by $14.18\%$ and $2.41\%$ while maintaining similar test accuracy.
EPIC-0.2 \cite{yang2022not} {(with $T=2, K=5$)} achieves the lowest poison success rate at $9.10\%$, but drops test accuracy to $86.21\%$, and \cite{gao2019strip} achieves $92.26\%$ test accuracy but suffers from $62.68\%$ poison accuracy. 

We also perform an ablation on each components of \alg\ in \tabref{tab:noise}. We show that naively applying Friendly Noise ($\zeta=32$) yields a high poison success rate of $10\%$. On the other hand, applying random noise ($\mu=32$) yields low poison success rates but also results in a significant test accuracy tradeoff (e.g. $>4.0\%$ drop for Gaussian and Bernoulli noise). Here, we show that applying \alg{} by proportionately combining friendly noise ($\zeta=16$) with each of the random noise components ($\mu=16$) maintains high test accuracy (i.e. only $2.0\%$ drop) while keeping poison success rate close to 0.

\subsubsection{Defending against Adaptive Attacks}

As discussed in \secref{sec:adaptive-attacks}, we believe that an adaptive attack against our defense is computationally prohibitive. To evaluate our claim, we modified the differentiable data augmentation component of Sleeper Agent attack algorithm to include randomly sampled Bernoulli noise. We then further added the fixed friendly noise generated from selected prior runs to the attacker’s augmentation procedure. Our results in \tabref{tab:adaptive} show that despite attacks being adapted to Bernoulli and friendly noise, poison accuracies are all within the standard deviation of one another, and the adaptive attacks cannot succeed.

\begin{table}[h]
    \centering
    \caption{Defense against Adaptive Attacks generated on Sleeper Agent averaged over 24 datasets (40 epoch setting). Bernoulli and friendly noise are both generated with $\epsilon=16$.}
    \label{tab:adaptive}
    \begin{tabular}{c|c|cc|cc}
    \toprule
        & & \multicolumn{2}{c}{Undefended} & \multicolumn{2}{c}{Defended (\alg-B)} \\
        Attack & Adaptation & Poison Acc & Test Acc & Poison Acc & Test Acc \\  
        \midrule
        Sleeper Agent & None & $83.48 \pm 7.58$ & $91.56 \pm 0.19$ & $21.52 \pm 8.39$ & $89.76 \pm 0.30$ \\ 
        Sleeper Agent & Bernoulli Noise & $80.25 \pm 8.13$ & $91.46 \pm 0.27$ & $31.05 \pm 9.45$ & $89.69 \pm 0.33$ \\ 
        Sleeper Agent & \alg-B & $79.08 \pm 8.30$ & $91.62 \pm 0.38$ & $30.50 \pm 9.40$ & $88.42 \pm 0.39$ \\
        \bottomrule
    \end{tabular}
\end{table}

\subsection{Transfer learning}
Next, we evaluate our method on data poisoning and backdoor attacks designed for the transfer learning scenario \cite{aghakhani2021bullseye,Shafahi2018poisonfrogs}. 
Here, the attacks are crafted based on a pretrained network with the goal of achieving poisoning when transfer learning is performed using the generated poisoned dataset. For the transfer learning scenario used in poisoning benchmarks \citep{geiping2021doesn,schwarzschild2020just}, the linear layer (classifier) of the pretrained model is re-initialized and trained with the poisoned dataset, while other layers (feature extractor) remain fixed during the training. Similar to the from-scratch setting, attacks are limited to a budget of 1\% and $\xi=16$. However, we generate \alg{} at the beginning of training instead of after 5 training epochs, since the feature extractor is already initialized. We note that this is not the true transfer learning setting, since the pretraining and transfer learning datasets are the same. However, as \citep{geiping2020witches,geiping2021doesn} noted, this presents an effective worst-case scenario to evaluate poisoning attacks. We show that in \tabref{tab:attacks} that even in such cases, we reduce poison success rate from $100\%$ to $35\%$ for the Bullseye Polytope attack \cite{aghakhani2021bullseye}, and from $100\%$ to $30\%$ for the Poison Frogs attack \cite{Shafahi2018poisonfrogs}.

\subsection{Transferability across Architectures}
\begin{table}[t]
    \centering
    \caption{Transferability between different architectures}\vspace{-1mm}
    \label{tab:architectures}
    \begin{tabular}{c|cc}
    \toprule
        Method & Poison Acc & Test Acc \\
        \toprule
         \alg-B (ResNet18) & 0\% & 91.52\% \\
         \alg-B (AlexNet -> ResNet18) & 0\% & 91.27\% \\ 
         \alg-B (LeNet -> ResNet18) & 0\% & 91.39\% \\
         \bottomrule
    \end{tabular}\vspace{-3mm}
\end{table}
We show that perturbations generated by \alg{} are transferable across architectures. In \tabref{tab:architectures}, we show using Gradient Matching $\xi=16$ that \alg{} optimized using smaller architectures, in particular AlexNet \citep{krizhevsky2012imagenet} and LeNet \citep{lecun98}, can be directly used for larger architectures like ResNet18. This presents a significant advantage in terms of computational costs, since \alg{} can be generated using smaller, and hence faster, models. Crucially, this makes the generated friendly noise free to be directly applied to (much larger) architectures.

\section{Conclusion}
We proposed a simple and highly effective defense mechanism, \alg, that protects deep learning pipelines against various types of poisoning attacks. 
Our defense is built on the observation that poisoning attacks 
introduce local sharp regions with high training loss, by adding adversarial perturbations to a subset of training examples.
\alg~relies on two components to break the poisons: an accuracy-friendly perturbation that is generated to maximally perturb examples without degrading the performance, and a randomly varying noise component.
The first component %
alleviates the local sharp regions introduced by poisons, and the second component further smooths out the loss landscape.
Both components combined together build a very light-weight but highly effective defense against the most powerful triggerless and backdoor poisoning attacks, including Gradient Matching, Bull-eyes Polytope, Poison Frogs, and Sleeper Agent, in transfer learning or training from scratch scenarios.
\alg~is extremely difficult to break with adaptive attacks and our friendly noise can be transferred to other architecture. This makes it almost free to apply to real-world deep learning pipelines.
Our defense is particularly targeted towards clean-label poisoning attacks that are generated under bounded perturbations. Such settings are the most difficult to defend, as generated poisons can easily fool even an expert observer. %
In contrast, unbounded attacks can be easily detected by manual or automated filtering mechanisms, through a single pass over the dataset.

\section{Acknowledgements}
This research was partially supported by Cisco Systems, the National Science Foundation CAREER Award 2146492, and the UCLA-Amazon Science Hub for Humanity and AI.

\bibliography{arxiv}

\begin{thebibliography}{10}

\bibitem{abadi2016deep}
Martin Abadi, Andy Chu, Ian Goodfellow, H~Brendan McMahan, Ilya Mironov, Kunal
  Talwar, and Li~Zhang.
\newblock Deep learning with differential privacy.
\newblock In {\em Proceedings of the 2016 ACM SIGSAC conference on computer and
  communications security}, pages 308--318, 2016.

\bibitem{aghakhani2021bullseye}
Hojjat Aghakhani, Dongyu Meng, Yu-Xiang Wang, Christopher Kruegel, and Giovanni
  Vigna.
\newblock Bullseye polytope: A scalable clean-label poisoning attack with
  improved transferability.
\newblock In {\em 2021 IEEE European Symposium on Security and Privacy
  (EuroS\&P)}, pages 159--178. IEEE, 2021.

\bibitem{berger2021stereoscopic}
Zachary Berger, Parth Agrawal, Tian~Yu Liu, Stefano Soatto, and Alex Wong.
\newblock Stereoscopic universal perturbations across different architectures
  and datasets.
\newblock {\em arXiv preprint arXiv:2112.06116}, 2021.

\bibitem{borgnia2021strong}
Eitan Borgnia, Valeriia Cherepanova, Liam Fowl, Amin Ghiasi, Jonas Geiping,
  Micah Goldblum, Tom Goldstein, and Arjun Gupta.
\newblock Strong data augmentation sanitizes poisoning and backdoor attacks
  without an accuracy tradeoff.
\newblock In {\em ICASSP 2021-2021 IEEE International Conference on Acoustics,
  Speech and Signal Processing (ICASSP)}, pages 3855--3859. IEEE, 2021.

\bibitem{borgnia2021dp}
Eitan Borgnia, Jonas Geiping, Valeriia Cherepanova, Liam Fowl, Arjun Gupta,
  Amin Ghiasi, Furong Huang, Micah Goldblum, and Tom Goldstein.
\newblock Dp-instahide: Provably defusing poisoning and backdoor attacks with
  differentially private data augmentations.
\newblock {\em arXiv preprint arXiv:2103.02079}, 2021.

\bibitem{cai2021learning}
Yuanhao Cai, Xiaowan Hu, Haoqian Wang, Yulun Zhang, Hanspeter Pfister, and
  Donglai Wei.
\newblock Learning to generate realistic noisy images via pixel-level
  noise-aware adversarial training.
\newblock {\em Advances in Neural Information Processing Systems}, 34, 2021.

\bibitem{chen2019detecting}
Bryant Chen, Wilka Carvalho, Nathalie Baracaldo, Heiko Ludwig, Benjamin
  Edwards, Taesung Lee, Ian Molloy, and Biplav Srivastava.
\newblock Detecting backdoor attacks on deep neural networks by activation
  clustering.
\newblock In {\em SafeAI@ AAAI}, 2019.

\bibitem{chen2017targeted}
Xinyun Chen, Chang Liu, Bo~Li, Kimberly Lu, and Dawn Song.
\newblock Targeted backdoor attacks on deep learning systems using data
  poisoning.
\newblock {\em arXiv preprint arXiv:1712.05526}, 2017.

\bibitem{gao2019strip}
Yansong Gao, Change Xu, Derui Wang, Shiping Chen, Damith~C. Ranasinghe, and
  Surya Nepal.
\newblock Strip: A defence against trojan attacks on deep neural networks.
\newblock In {\em Proceedings of the 35th Annual Computer Security Applications
  Conference}, ACSAC '19, page 113–125, New York, NY, USA, 2019. Association
  for Computing Machinery.

\bibitem{geiping2020witches}
Jonas Geiping, Liam Fowl, W~Ronny Huang, Wojciech Czaja, Gavin Taylor, Michael
  Moeller, and Tom Goldstein.
\newblock Witches' brew: Industrial scale data poisoning via gradient matching.
\newblock {\em arXiv preprint arXiv:2009.02276}, 2020.

\bibitem{geiping2021doesn}
Jonas Geiping, Liam Fowl, Gowthami Somepalli, Micah Goldblum, Michael Moeller,
  and Tom Goldstein.
\newblock What doesn't kill you makes you robust (er): Adversarial training
  against poisons and backdoors.
\newblock {\em arXiv preprint arXiv:2102.13624}, 2021.

\bibitem{geiping2021witches}
Jonas Geiping, Liam~H Fowl, W.~Ronny Huang, Wojciech Czaja, Gavin Taylor,
  Michael Moeller, and Tom Goldstein.
\newblock Witches' brew: Industrial scale data poisoning via gradient matching.
\newblock In {\em International Conference on Learning Representations}, 2021.

\bibitem{gu2017badnets}
Tianyu Gu, Brendan Dolan-Gavitt, and Siddharth Garg.
\newblock Badnets: Identifying vulnerabilities in the machine learning model
  supply chain.
\newblock {\em arXiv preprint arXiv:1708.06733}, 2017.

\bibitem{hong2020effectiveness}
Sanghyun Hong, Varun Chandrasekaran, Yi{\u{g}}itcan Kaya, Tudor Dumitra{\c{s}},
  and Nicolas Papernot.
\newblock On the effectiveness of mitigating data poisoning attacks with
  gradient shaping.
\newblock {\em arXiv preprint arXiv:2002.11497}, 2020.

\bibitem{huang2020metapoison}
W~Ronny Huang, Jonas Geiping, Liam Fowl, Gavin Taylor, and Tom Goldstein.
\newblock Metapoison: Practical general-purpose clean-label data poisoning.
\newblock {\em Advances in Neural Information Processing Systems}, 33, 2020.

\bibitem{jayaraman2019evaluating}
Bargav Jayaraman and David Evans.
\newblock Evaluating differentially private machine learning in practice.
\newblock In {\em 28th $\{$USENIX$\}$ Security Symposium ($\{$USENIX$\}$
  Security 19)}, pages 1895--1912, 2019.

\bibitem{koh2018stronger}
Pang~Wei Koh, Jacob Steinhardt, and Percy Liang.
\newblock Stronger data poisoning attacks break data sanitization defenses.
\newblock {\em arXiv preprint arXiv:1811.00741}, 2018.

\bibitem{krizhevsky2012imagenet}
Alex Krizhevsky, Ilya Sutskever, and Geoffrey~E Hinton.
\newblock Imagenet classification with deep convolutional neural networks.
\newblock In F.~Pereira, C.~J.~C. Burges, L.~Bottou, and K.~Q. Weinberger,
  editors, {\em Advances in Neural Information Processing Systems 25}, pages
  1097--1105. Curran Associates, Inc., 2012.

\bibitem{ShankarAML}
Ram Shankar~Siva Kumar, Magnus Nyström, John Lambert, Andrew Marshall, Mario
  Goertzel, Andi Comissoneru, Matt Swann, and Sharon Xia.
\newblock Adversarial machine learning -- industry perspectives, 2020.

\bibitem{lecun98}
Y.~Lecun, L.~Bottou, Y.~Bengio, and P.~Haffner.
\newblock Gradient-based learning applied to document recognition.
\newblock {\em Proceedings of the IEEE}, 86(11):2278--2324, 1998.

\bibitem{levine2020deep}
Alexander Levine and Soheil Feizi.
\newblock Deep partition aggregation: Provable defenses against general
  poisoning attacks.
\newblock In {\em International Conference on Learning Representations}, 2020.

\bibitem{li2021anti}
Yige Li, Xixiang Lyu, Nodens Koren, Lingjuan Lyu, Bo~Li, and Xingjun Ma.
\newblock Anti-backdoor learning: Training clean models on poisoned data.
\newblock {\em Advances in Neural Information Processing Systems}, 34, 2021.

\bibitem{liu2017trojaning}
Yingqi Liu, Shiqing Ma, Yousra Aafer, Wen-Chuan Lee, Juan Zhai, Weihang Wang,
  and Xiangyu Zhang.
\newblock Trojaning attack on neural networks.
\newblock {\em Purdue University Department of Computer Science Technical
  Reports}, 2017.

\bibitem{madaan2021learning}
Divyam Madaan, Jinwoo Shin, and Sung~Ju Hwang.
\newblock Learning to generate noise for multi-attack robustness.
\newblock In {\em International Conference on Machine Learning}, pages
  7279--7289. PMLR, 2021.

\bibitem{madry2018towards}
Aleksander Madry, Aleksandar Makelov, Ludwig Schmidt, Dimitris Tsipras, and
  Adrian Vladu.
\newblock Towards deep learning models resistant to adversarial attacks.
\newblock In {\em International Conference on Learning Representations}, 2018.

\bibitem{naseer2019cross}
Muhammad~Muzammal Naseer, Salman~H Khan, Muhammad~Haris Khan, Fahad
  Shahbaz~Khan, and Fatih Porikli.
\newblock Cross-domain transferability of adversarial perturbations.
\newblock {\em Advances in Neural Information Processing Systems}, 32, 2019.

\bibitem{peri2020deep}
Neehar Peri, Neal Gupta, W~Ronny Huang, Liam Fowl, Chen Zhu, Soheil Feizi, Tom
  Goldstein, and John~P Dickerson.
\newblock Deep k-nn defense against clean-label data poisoning attacks.
\newblock In {\em European Conference on Computer Vision}, pages 55--70.
  Springer, 2020.

\bibitem{qin2021random}
Zeyu Qin, Yanbo Fan, Hongyuan Zha, and Baoyuan Wu.
\newblock Random noise defense against query-based black-box attacks.
\newblock {\em Advances in Neural Information Processing Systems}, 34, 2021.

\bibitem{rusak2020simple}
Evgenia Rusak, Lukas Schott, Roland~S Zimmermann, Julian Bitterwolf, Oliver
  Bringmann, Matthias Bethge, and Wieland Brendel.
\newblock A simple way to make neural networks robust against diverse image
  corruptions.
\newblock In {\em European Conference on Computer Vision}, pages 53--69.
  Springer, 2020.

\bibitem{Saha2019htbd}
Aniruddha Saha, Akshayvarun Subramanya, and Hamed Pirsiavash.
\newblock Hidden trigger backdoor attacks, 2019.

\bibitem{schwarzschild2020just}
Avi Schwarzschild, Micah Goldblum, Arjun Gupta, John~P Dickerson, and Tom
  Goldstein.
\newblock Just how toxic is data poisoning? a unified benchmark for backdoor
  and data poisoning attacks.
\newblock {\em arXiv preprint arXiv:2006.12557}, 2020.

\bibitem{Shafahi2018poisonfrogs}
Ali Shafahi, W.~Ronny Huang, Mahyar Najibi, Octavian Suciu, Christoph Studer,
  Tudor Dumitras, and Tom Goldstein.
\newblock Poison frogs! targeted clean-label poisoning attacks on neural
  networks, 2018.

\bibitem{shokri2020bypassing}
Reza Shokri et~al.
\newblock Bypassing backdoor detection algorithms in deep learning.
\newblock In {\em 2020 IEEE European Symposium on Security and Privacy
  (EuroS\&P)}, pages 175--183. IEEE, 2020.

\bibitem{souri2021sleeper}
Hossein Souri, Micah Goldblum, Liam Fowl, Rama Chellappa, and Tom Goldstein.
\newblock Sleeper agent: Scalable hidden trigger backdoors for neural networks
  trained from scratch.
\newblock {\em arXiv preprint arXiv:2106.08970}, 2021.

\bibitem{Steinhardt17certified}
Jacob Steinhardt, Pang~Wei Koh, and Percy Liang.
\newblock Certified defenses for data poisoning attacks, 2017.

\bibitem{szegedy2013intriguing}
Christian Szegedy, Wojciech Zaremba, Ilya Sutskever, Joan Bruna, Dumitru Erhan,
  Ian Goodfellow, and Rob Fergus.
\newblock Intriguing properties of neural networks.
\newblock {\em arXiv preprint arXiv:1312.6199}, 2013.

\bibitem{tao2021better}
Lue Tao, Lei Feng, Jinfeng Yi, Sheng-Jun Huang, and Songcan Chen.
\newblock Better safe than sorry: Preventing delusive adversaries with
  adversarial training.
\newblock {\em Advances in Neural Information Processing Systems}, 34, 2021.

\bibitem{tran2018spectral}
Brandon Tran, Jerry Li, and Aleksander Madry.
\newblock Spectral signatures in backdoor attacks.
\newblock In {\em Advances in Neural Information Processing Systems}, pages
  8000--8010, 2018.

\bibitem{turner2018clean}
Alexander Turner, Dimitris Tsipras, and Aleksander Madry.
\newblock Clean-label backdoor attacks.
\newblock {\em OpenReview}, 2018.

\bibitem{wang2019neural}
Bolun Wang, Yuanshun Yao, Shawn Shan, Huiying Li, Bimal Viswanath, Haitao
  Zheng, and Ben~Y Zhao.
\newblock Neural cleanse: Identifying and mitigating backdoor attacks in neural
  networks.
\newblock In {\em 2019 IEEE Symposium on Security and Privacy (SP)}, pages
  707--723. IEEE, 2019.

\bibitem{weber2020rab}
Maurice Weber, Xiaojun Xu, Bojan Karla{\v{s}}, Ce~Zhang, and Bo~Li.
\newblock Rab: Provable robustness against backdoor attacks.
\newblock {\em arXiv preprint arXiv:2003.08904}, 2020.

\bibitem{xie2019improving}
Cihang Xie, Zhishuai Zhang, Yuyin Zhou, Song Bai, Jianyu Wang, Zhou Ren, and
  Alan~L Yuille.
\newblock Improving transferability of adversarial examples with input
  diversity.
\newblock In {\em Proceedings of the IEEE/CVF Conference on Computer Vision and
  Pattern Recognition}, pages 2730--2739, 2019.

\bibitem{yang2022not}
Yu~Yang, Tian~Yu Liu, and Baharan Mirzasoleiman.
\newblock Not all poisons are created equal: Robust training against data
  poisoning.
\newblock In {\em International Conference on Machine Learning}, pages
  25154--25165. PMLR, 2022.

\bibitem{zhu2019transferable}
Chen Zhu, W~Ronny Huang, Hengduo Li, Gavin Taylor, Christoph Studer, and Tom
  Goldstein.
\newblock Transferable clean-label poisoning attacks on deep neural nets.
\newblock In {\em International Conference on Machine Learning}, pages
  7614--7623, 2019.

\end{thebibliography}
\bibliographystyle{plain}

\newpage
\appendix

\section*{\begin{center}\Large{Supplementary Material:\\ Friendly Noise against Adversarial Noise:
A Powerful Defense against Data Poisoning Attacks}\end{center}}

\section{Additional Experiments}
\subsection{Ablation on different norms for friendly noise objective}
\label{app:norm-ablation}
Recall the friendly noise objective is given by the following
\begin{align}
    \epsilon_i = \argmin_{\epsilon:\|\epsilon\|_\infty\leq\zeta} {D_{KL}\big(f_\theta(x_i+\epsilon) || f_\theta(x_i)\big)} - \lambda {\|\epsilon\|}_2.
\end{align}
We note that while we used the $L_2$ norm for encouraging larger values of $\epsilon$, other norms such as $L_{\infty}$ and $L_1$ can also be used. $L_{\infty}$ only cares about the largest element in the noise vector, so it adds larger perturbations, while $L_1$ cares about absolute values hence it adds smaller (but more non-zero) perturbations. 
$L_2$ encourages larger noise elements compared to $L_1$. \tabref{tab:norms-ablation} shows that different norms result in similar poison and test accuracy. 
\begin{table}[H]
    \centering
    \caption{Ablation of different norms for the friendly noise objective on CIFAR-10 - Gradient Matching ($\xi$ = 16) Performance is similar across L1/L2 norms. Here we use FRIENDs-B with the same defense settings as \tabref{tab:wb16}.}
    \begin{tabular}{ccc}
        \toprule
         Norm & Poison Acc & Test Acc  \\
         \midrule
         $L_2$ & 0.00\% & 91.52\% \\
         $L_1$ & 0.00\% & 91.50\% \\
         $L_\infty$ & 0.00\% & 91.37\% \\
         \bottomrule
    \end{tabular}
    \label{tab:norms-ablation}
\end{table}

\section{Additional Visualizations}
\subsection{Histogram visualizations of friendly noise and its variants}
\label{app:histogram-friends}

\begin{figure}[H]
\centering
\begin{subfigure}[Friendly Noise\label{fig:histogram-friendly-noise}]{
\includegraphics[width=.24\textwidth]{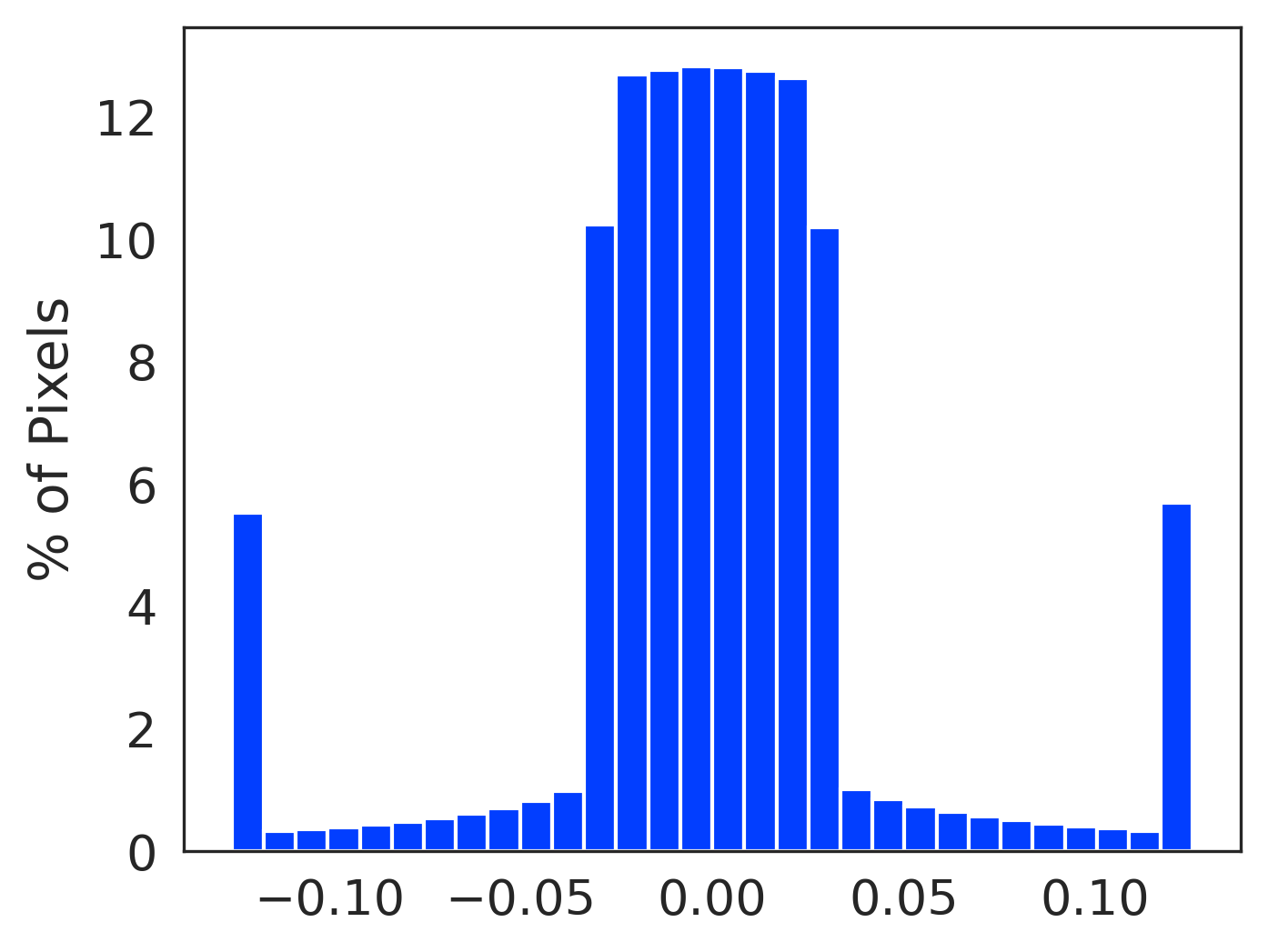}
}\hspace{-3mm}
\end{subfigure}
\begin{subfigure}[FrieNDs-U\label{fig:histogram-friends-u}]{
\includegraphics[width=.24\textwidth]{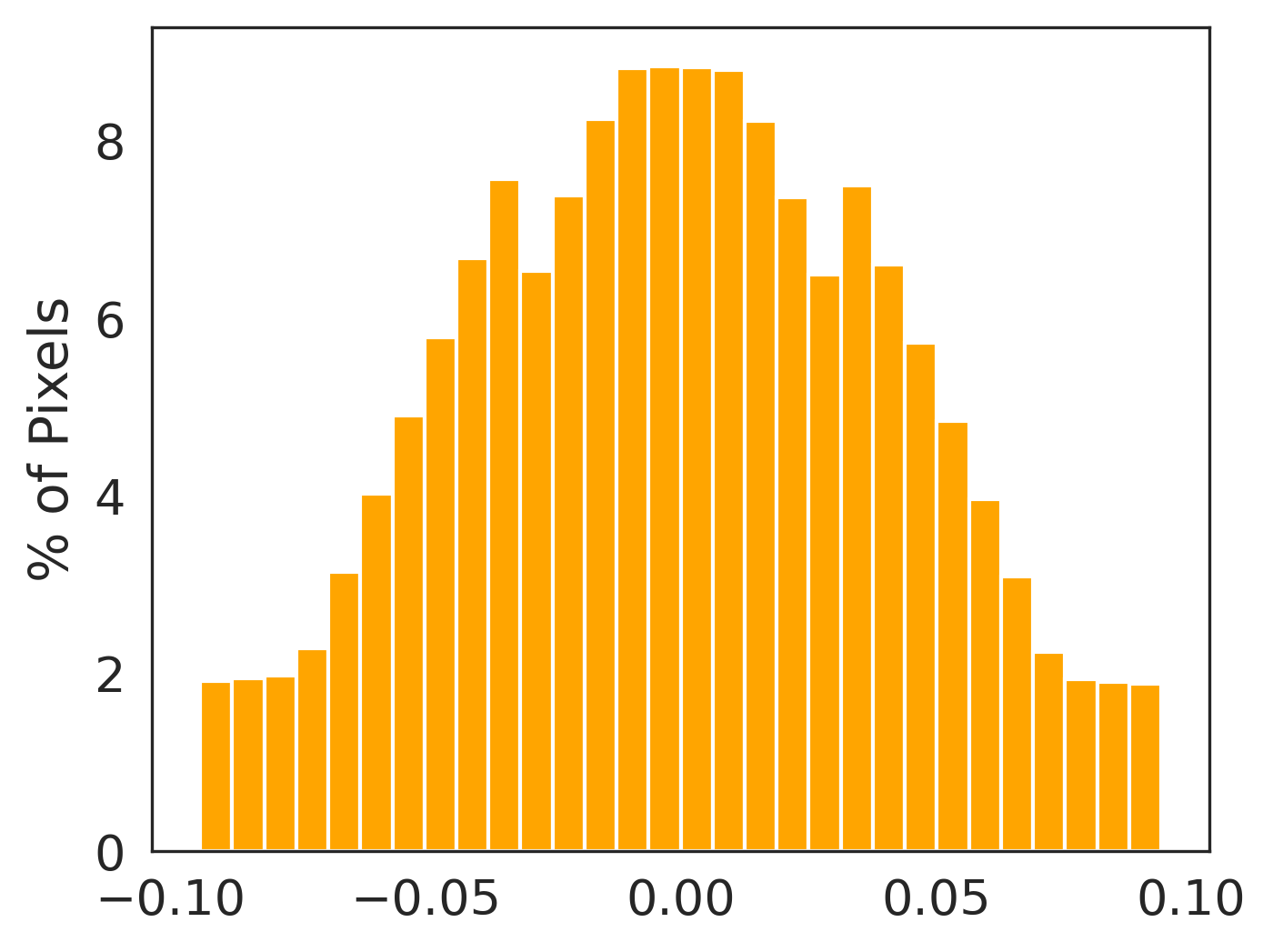}
}\hspace{-3mm}
\end{subfigure}
\begin{subfigure}[FrieNDs-G\label{fig:histogram-friends-g}]{
\includegraphics[width=.235\textwidth]{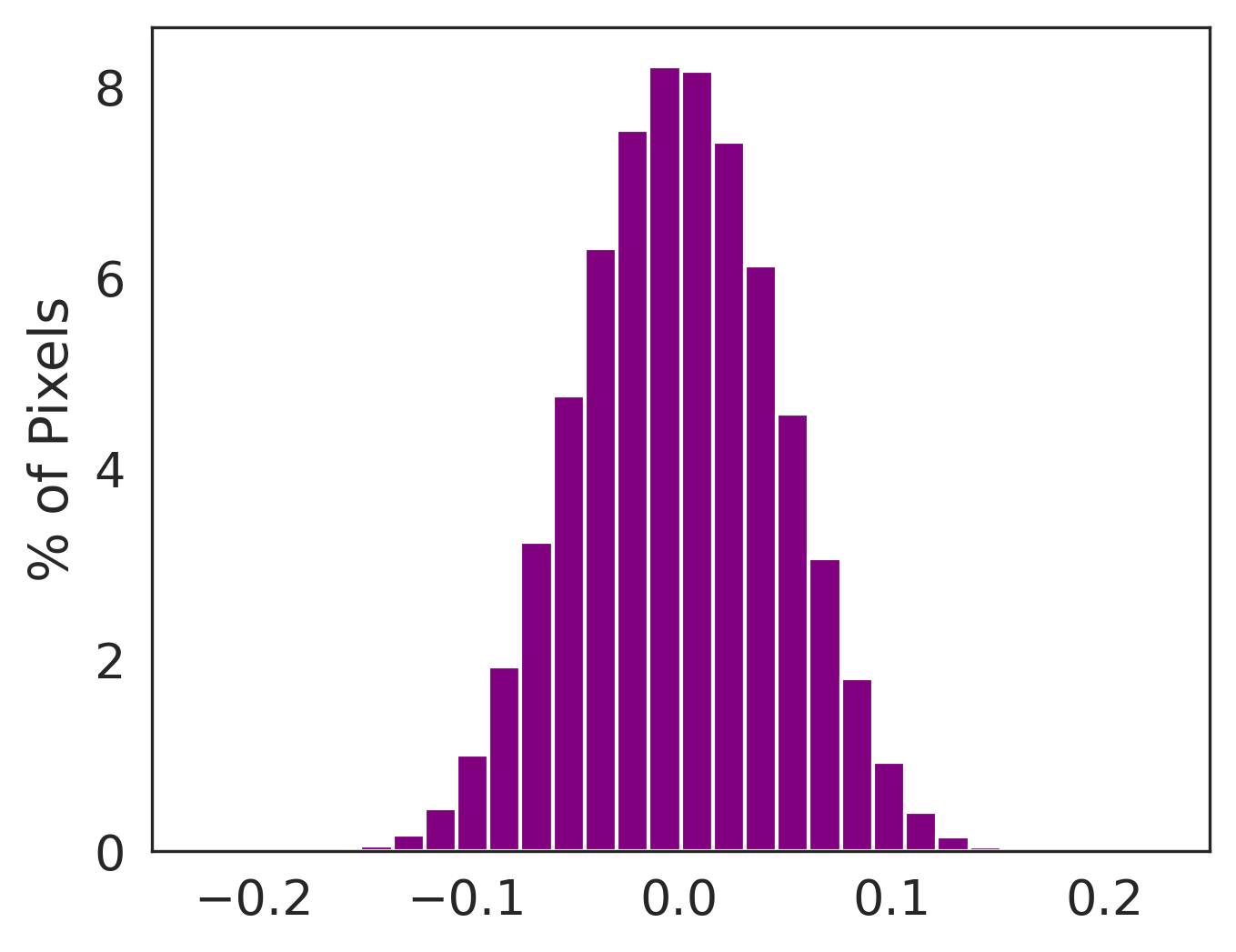}
}\hspace{-3mm}
\end{subfigure}
\begin{subfigure}[FrieNDs-B\label{fig:histogram-friends-b}]{
\includegraphics[width=.25\textwidth]{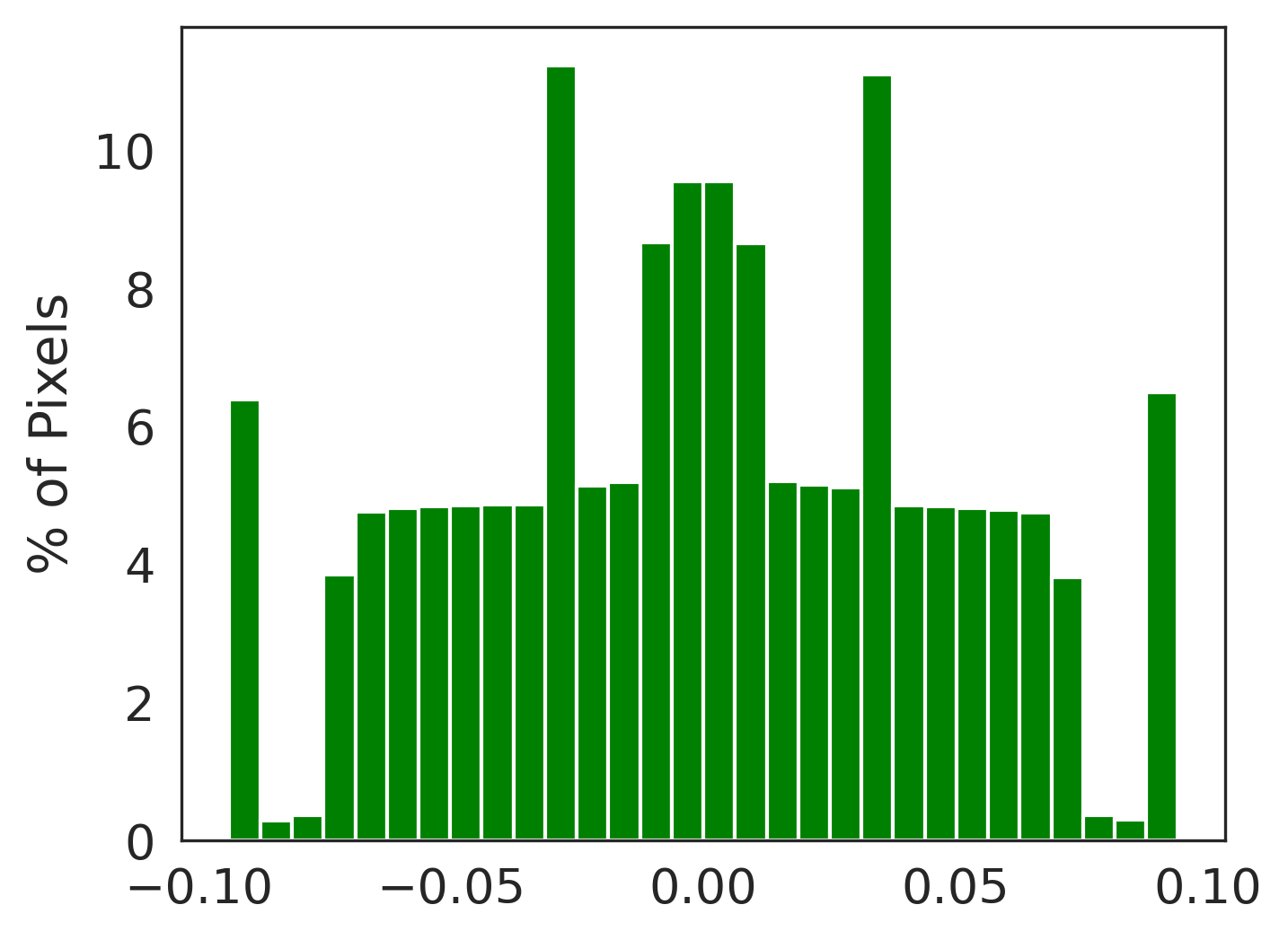}
}
\end{subfigure}
\caption{Histogram of our method with different types of random noises For (a), we set $\zeta=32$. For (b)-(d), we set $\zeta=16,\mu=16$.}
\label{fig:histogram-friends-all}
\end{figure}

 \figref{fig:histogram-friends-all} compares the distribution of random noise sampled from these distributions combined with the optimized perturbation obtained from Eq. \eqref{eqn:noise-objective}. We can see that uniform noise perturbs all the pixels similarly, and hence small amount of uniform noise does not harm the model's performance but cannot effectively breaks poisons. Larger uniform noise, however, has a larger effect on the model's performance.
Bernoulli and Gaussian noise on the other hand add a larger perturbation to individual pixels.
Hence, they are more effective in reducing the attack success rate, but they harm the test accuracy more as they the larger perturbation may be added to the more sensitive areas.
\Cref{fig:histogram-friends-u,fig:histogram-friends-g,fig:histogram-friends-b} shows the distribution of random noise combined with our optimized noise added to different pixels. %
{We observe that random noise is added to regions where friendly noise is less dominant, hence resulting in a significant perturbation to an overall greater number of pixels to break poisoning attacks.}

\subsection{Trade-off between the time of generating friendly noise and their effectiveness}
\label{app:friends-begin-time-tradeoff}
\begin{figure}[H]
\centering
\begin{subfigure}[\label{fig:poison_epoch}]{
\includegraphics[width=.305\textwidth]{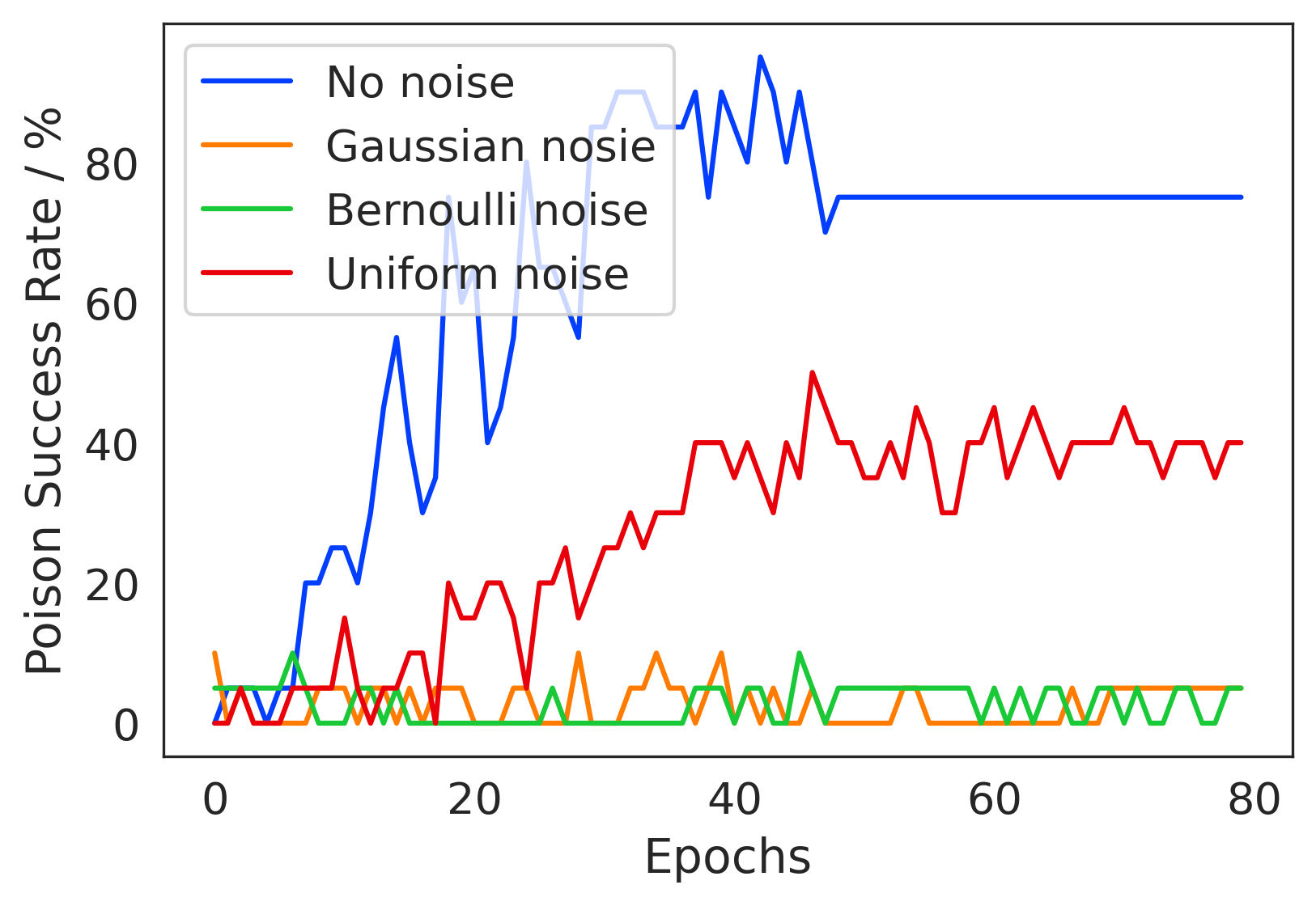}
\vspace{-1mm}
}%
\end{subfigure}
\begin{subfigure}[\label{fig:acc_epoch}]{
\includegraphics[width=.305\textwidth]{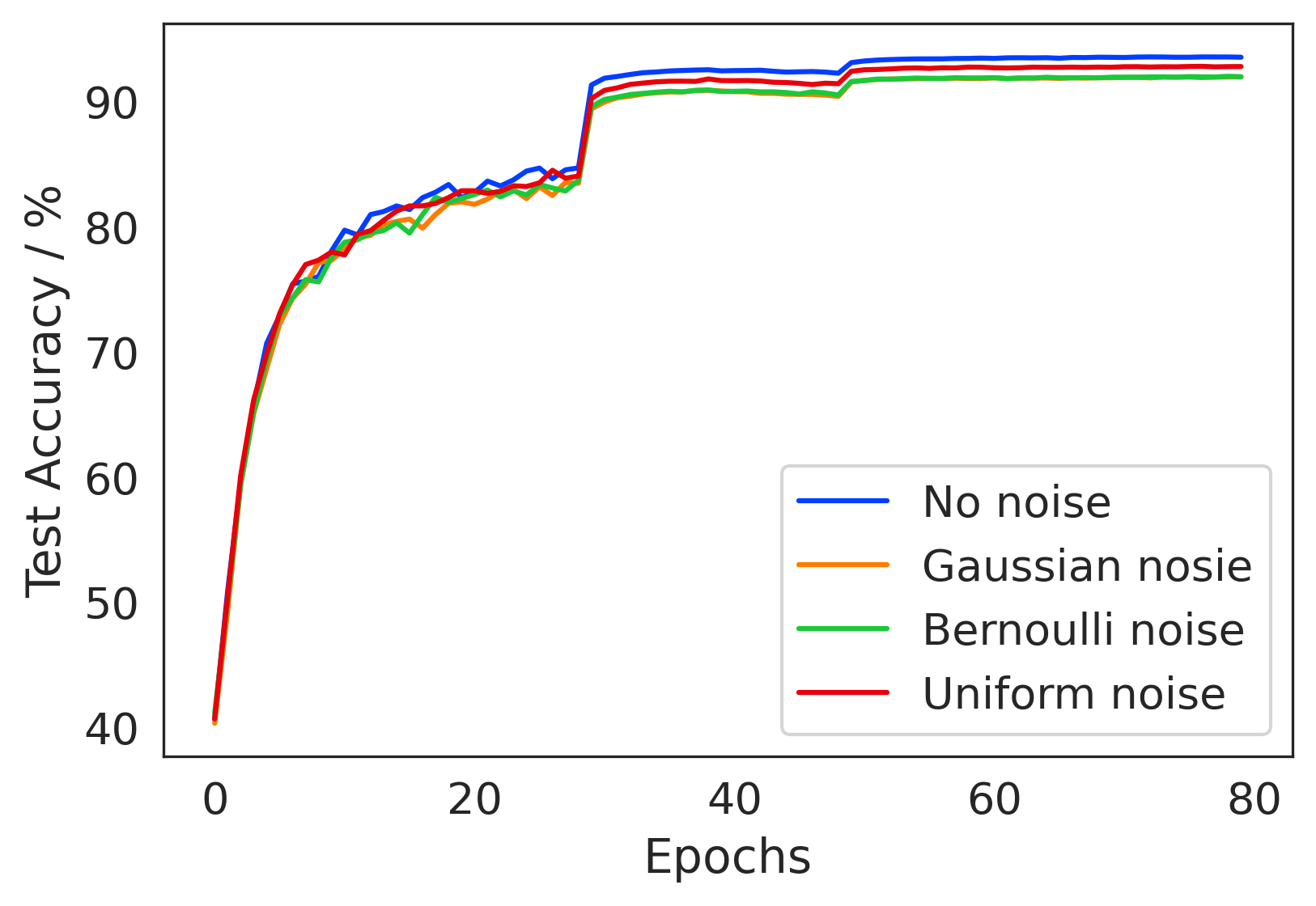}
\vspace{-1mm}
}
\end{subfigure}
\begin{subfigure}[\label{fig:noise_deviations}]{
 \includegraphics[width=.325\textwidth]{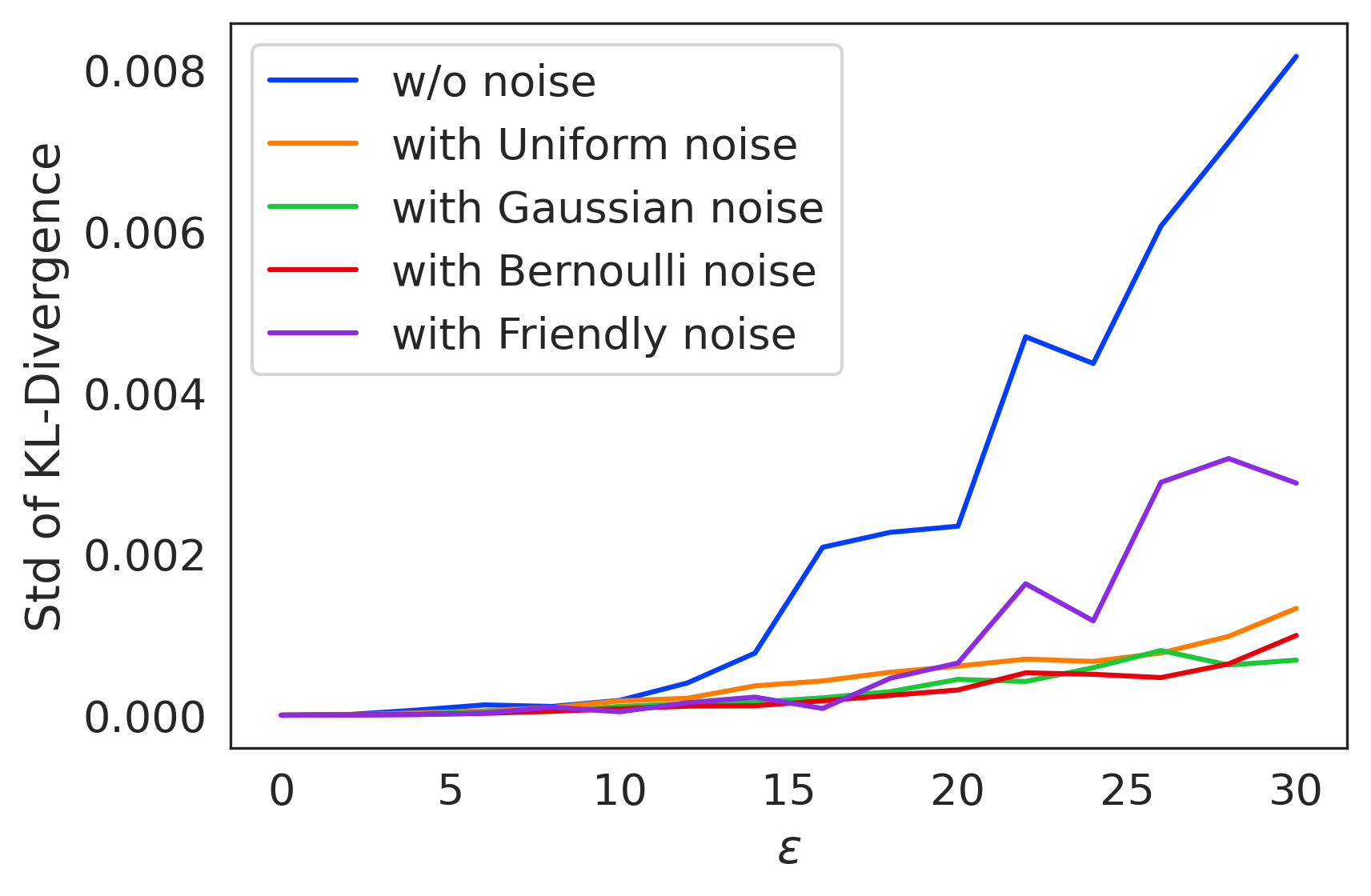}
 \vspace{-1mm}
}
\end{subfigure}
\caption{(a) Number of poisoned datasets vs epochs. It takes at least a few epochs for the poisons to have an effect. (b) On the other hand, generalization of the predictions and features learnt increases over time, as measured by error on the test set. %
(c) %
Standard deviation of KL-divergence (y-axis) between predictions of training examples before and after adding friendly perturbations in Eq. \eqref{eqn:noise-objective} and random noise sampled from various distributions. Standard deviation is calculated over 10 randomly sampled points in an $\xi$ balls (x-axis) around every training example.}
\end{figure}

\end{document}